\documentclass[10pt]{article}

\usepackage{fullpage}

\usepackage{amsmath}
\usepackage{graphicx}
\usepackage{amssymb}
\usepackage{subfigure}
\usepackage{color}

\usepackage{fullpage}

\newcommand\Ro[1][\relax]{\ifx\relax#1 \ensuremath{\mathcal{R}_0}
  \else \ensuremath{\mathcal{R}_{0,#1}} \fi}
\newcommand{\I}{\mathcal{I}}
\newcommand{\Sus}{\mathcal{S}}
\newcommand{\PE}{\mathcal{P}}
\newcommand{\A}{\mathcal{A}}
\newcommand{\order}{\mathcal{O}}
\newcommand{\ave}[1]{\left \langle #1 \right \rangle}
\newcommand{\E}{\mathbb{E}}
\newcommand{\erdosrenyi}{Erd\H{o}s--R\'{e}nyi}

\newcommand{\di}{\raisebox{-2pt}[0pt][0pt]{\includegraphics{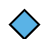}}}
\newcommand{\sq}{\raisebox{-2pt}[0pt][0pt]{\includegraphics{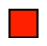}}}
\newcommand{\ex}{\raisebox{-2pt}[0pt][0pt]{\includegraphics{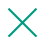}}}
\newcommand{\ci}{\raisebox{-2pt}[0pt][0pt]{\includegraphics{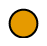}}}
\newcommand{\pent}{\raisebox{-2pt}[0pt][0pt]{\includegraphics{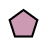}}}

\newcommand{\pe}{\hat{p}}

\title{Spread of infectious disease through clustered populations}

\author{Joel C. Miller}
\date{\today}

\begin{document}

\maketitle

\begin{abstract}
  Networks of person-person contacts form the substrate along which
  infectious diseases spread.  Most network-based studies of the
  spread focus on the impact of variations in degree (the number of
  contacts an individual has).  However, other effects such as
  clustering, variations in infectiousness or susceptibility, or
  variations in closeness of contacts may play a significant role.  We
  develop analytic techniques to predict how these effects alter the
  growth rate, probability, and size of epidemics and validate the
  predictions with a realistic social network.  We find that (for
  given degree distribution and average transmissibility) clustering
  is the dominant factor controlling the growth rate, heterogeneity in
  infectiousness is the dominant factor controlling the probability of
  an epidemic, and heterogeneity in susceptibility is the dominant
  factor controlling the size of an epidemic.  Edge weights (measuring
  closeness or duration of contacts) have impact only if correlations
  exist between different edges.  Combined, these effects can play a
  minor role in reinforcing one another, with the impact of clustering
  largest when the population is maximally heterogeneous or if the
  closer contacts are also strongly clustered.  Our most significant
  contribution is a systematic way to address clustering in infectious
  disease models, and our results have a number of implications for
  the design of interventions.
\end{abstract}

\section{Introduction}
\label{sec:introduction}

Recently H5N1 avian influenza and SARS have raised the profile of
emerging infectious diseases.  Both can infect humans, but have a
primary animal host.  Typically such zoonotic diseases emerge
periodically into the human population and disappear (\emph{e.g.},
Ebola, Hanta Virus, and Rabies), but sometimes (\emph{e.g.}, HIV) the
disease achieves sustained person-to-person spread.  With the advent
of modern transportation networks, diseases that formerly emerged in
isolated villages and died out without further spread may now spread
worldwide.

A number of interventions are available to control emerging diseases,
each with distinct costs and benefits.  To design optimal policies, we
must address several related, but nevertheless distinct, questions.
How fast would an epidemic spread?  How likely is a single
introduced infection to result in an epidemic?  How many people would an
epidemic infect?  We quantify these using $\Ro$, the \emph{basic
  reproductive ratio}, which measures the average number of new cases
each infection causes early in the outbreak; $\PE$, the probability
that a single infection sparks an epidemic; and $\A$, the
\emph{attack rate} or fraction of the population infected in an
epidemic.  Understanding these different quantities and what affects
them helps us to select policies with maximal impact for given cost.

Many different models are used to study disease spread.  Perhaps the
most important decision in developing a model is how the interactions
of the population are represented.  Because of the complexity of the
population, it is invariably necessary to make simplifying
assumptions.  The errors (and therefore, the conclusions) resulting
from many of these approximations are not well-quantified.  In this
paper we will focus on quantifying the impact of clustering (the
tendency to interact in small groups) and individual-scale
heterogeneity on the spread of an epidemic.

Based on how they handle clustering, models for population structure
fit into a hierarchy of three classes (which in turn may be
subdivided).  At the simplest level the population is assumed to mix
without any clustering.  Most existing models fall into this category.
At the most complex level, agent-based models are used: the movements
of each individual are tracked, and people who are in the same
location are able to infect one another.  These models typically
require significant resources to develop, and the clustering is
explicitly included.  An intermediate level of complexity attempts to
introduce the clustering as a parameter (or several parameters).
Usually these models only consider clustering in terms of the number
of triangles in a network, but as we shall see, other structures may
play a role.

Before introducing the details of our model, we review some previous
work.  All the models we consider are Susceptible-Infected-Recovered
(SIR) epidemic models~\cite{andersonmay}, in which individuals begin
susceptible, become infected by contacting infected individuals, and
finally recover with immunity.

For unclustered populations, ordinary differential equation (ODE)
models were among the earliest models used~\cite{kermack} and remain
the most common.  They are deterministic, and so cannot directly
calculate $\PE$, but they give insight into the factors controlling
$\Ro$ and $\A$.  Because they assume mass-action mixing, it is
difficult to incorporate individual heterogeneity in the number of
contacts.  More recently some network-based models have been
introduced for unclustered
populations~\cite{andersson:limit,newman:spread,kenah:second,miller:heterogeneity,meyers:contact,meyers:directed,meyers:sars}.
These models represent the population as nodes with edges between
nodes representing contacts, along which disease spreads
stochastically.  Heterogeneity in the number of contacts is introduced
by modifying the degree (number of edges) of each node.  By neglecting
clustering, these studies are able to make analytic predictions
through branching process arguments.  A recent sociological
study~\cite{mossong:PLOScontacts} used surveys with participants
recording the length and nature of their contacts.  This data is
valuable for providing the contact distribution needed for the above
network models, and allows us to apply network results to real
populations.  However, this data does not directly tell us anything
about the clustering of the population resulting from
family/work/other groups.  Other recent work
by~\cite{miller:heterogeneity,kenah:second} analytically addresses the
impact of heterogeneity in infectiousness and susceptibility in
unclustered networks.

Using agent-based
simulations~\cite{eubanks:episims,barrett,valle:episims,ferguson:SEAsia,germann:epicast,ajelli}
allows us to directly incorporate clustering.  In these simulations, the
population is a collection of individuals who move and contact one
another.  The modeller has complete control over the parameters
governing interactions and how the disease spreads.  This allows us to
study many effects, but also introduces many parameters.  It is difficult
to test the accuracy of the assumptions used to generate these models
and to extract which parameters are essential to the disease dynamics.
The expense of developing these simulations is frequently prohibitive.

In this paper we introduce a systematic approach for calculating the
impact of clustering, and quantifying the error.  Because our model
investigates disease spread in clustered networks, we provide a more
detailed review of previous work on clustering and disease.  A few
investigations have been made into the interaction of clustering with
disease spread using network models.  The attempts that have been
made~\cite{keeling:local,eames:clustered,newman:clustering,serrano:2,serrano:prl,britton:cluster}
typically use approximations whose errors are not quantified,
resulting in apparently contradictory results.  A few
papers~\cite{miller:bounds,trapman,kuulasmaa:locallydep} have
considered clustering and heterogeneities, rigorously showing that
increased heterogeneity tends to decrease $\PE$ and $\A$, but without
quantitative predictions.  Recently~\cite{eames:clustered} considered
the spread of epidemics in a class of random networks for which the
number of triangles could be controlled.  It may be inferred from
their figure 3 that clustering decreases the growth rate and that
sufficient clustering can increase the epidemic threshold.  However,
at small and moderate levels clustering appears not to alter the final
size of epidemics significantly.  Similar observations have been made
by~\cite{bansal:thesis}.  At first glance, this contradicts
observations of~\cite{serrano:2,serrano:prl} that clustering
significantly reduces the size of epidemics, but that sufficiently
strong clustering reduces the epidemic threshold (see
also~\cite{newman:clustering}), allowing epidemics at lower
transmissibility.  The discrepancy in epidemic size may be resolved by
noting that the networks in~\cite{serrano:2,serrano:prl} have low
average degree.  We will see that
clustering only affects the size if the typical degree is small or
clustering is very high.  The apparent discrepancy in epidemic
threshold with strong clustering may be resolved by noting that the
form of strong clustering considered by~\cite{serrano:2,serrano:prl}
forces preferential contacts between high degree nodes.  The reduction
in epidemic threshold is perhaps better understood in terms of
degree-degree correlations than in terms of clustering.

In this paper we develop techniques to incorporate general small-scale
structure (beyond triangles) into the calculation of $\Ro$, $\PE$, and
$\A$.  To calculate $\Ro$, we develop a systematic series expansion
which allows us to interpolate between unclustered and clustered
results by including more terms.  To calculate $\PE$ and $\A$, we use
a similar approach, but only give estimates on the size of correction
terms.  Our methods give us a rigorous means to understand how
unclustered results relate to more realistic populations, and our
results resolve the apparent discrepancies mentioned above.  Our
theory accurately predicts epidemic behaviour in a more realistic
contact network derived from an agent-based simulation of Portland,
Oregon by EpiSimS~\cite{valle:episims}.  We expand this to investigate
the interplay of clustering, heterogeneities in individual infectiousness or
susceptibility, and variation in edge weights in their
effect on $\Ro$, $\PE$, and $\A$.

The paper is organised as follows: Section~\ref{sec:formulation}
describes our model and networks and summarises earlier work on
unclustered networks.  These results will be the leading order terms
for our expansions for clustered networks in the remainder of the
paper.  Section~\ref{sec:homogeneous} considers how epidemics spread
in a clustered network assuming homogeneous transmission.  We derive
the corrections to $\Ro$ and show that the corrections to $\PE$ and
$\A$ are insignificant unless the typical degree is small or
clustering very high.  Section~\ref{sec:heterogeneous} considers
epidemics in clustered networks with heterogeneous infectiousness or
susceptibility, building on section~\ref{sec:homogeneous}.
Section~\ref{sec:weighted} extends this further to consider epidemics
spreading on clustered networks with weighted edges.  Edges with large
weights tend to occur in family or work groups which magnifies the
impact of clustering.  Finally section~\ref{sec:discussion} discusses
the implications of our results, particularly for designing
interventions.  We conclude that in general, heterogeneity
significantly impacts $\PE$ and $\A$, but not $\Ro$, while clustering
impacts $\Ro$ significantly, but not $\PE$ and $\A$.  Heterogeneity or
edge weights may enhance the impact of clustering.

\section{Formulation}
\label{sec:formulation}

\begin{figure}
\includegraphics[width = 0.3\columnwidth]{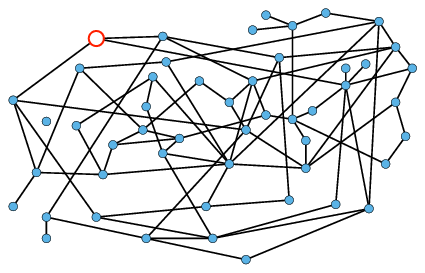}\hfill
\includegraphics[width = 0.3\columnwidth]{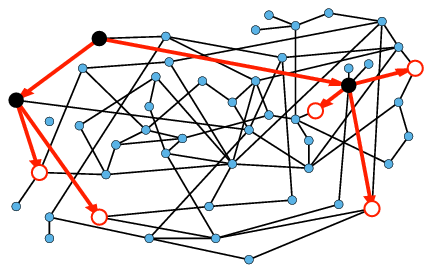}\hfill
\includegraphics[width = 0.3\columnwidth]{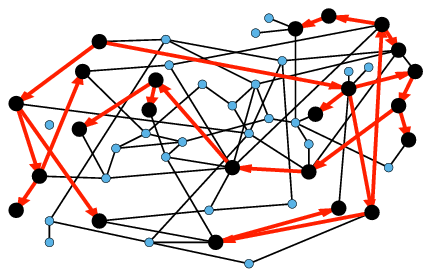}
\caption{A sample network and several stages of an outbreak.  Nodes
  begin susceptible (small circles), become infected (empty large
  circles), possibly infecting others along edges, and then recover
  (solid large circles).  The outbreak finishes when no infected nodes
  remain.}
\label{fig:sample}
\end{figure}

\subsection{The disease model}
We consider the spread of a disease using a discrete SIR model on a
static network $G$.  Nodes of $G$ represent individuals and edges
represent (potentially infectious) contacts.  The contact structure of
the network is fixed during the course of the outbreak.  The
\emph{degree} $k$ of a node $u$ is the number of edges containing $u$.
Figure~\ref{fig:sample} shows a sample outbreak.  A single infection, the
\emph{index case} is chosen uniformly from the population to begin an
\emph{outbreak}.  Infection spreads along an edge from an infected
node $u$ to a susceptible node $v$ with probability $T_{uv}$, the
\emph{transmissibility}.  The time it takes for infection and recovery
to occur may vary but  does not affect our results.  Once $u$
recovers it cannot be reinfected.  Typically for a large random
network with a population of $N=|G|$ nodes, the final size of
outbreaks is either large, with $\order(N)$ cumulative infections, or
small, with $\order(\log N)$ infections~\cite{bollobas:random}.  Large
outbreaks are \emph{epidemics} and small outbreaks are
\emph{non-epidemic outbreaks}.

\subsubsection{Transmissibility}
A number of factors influence the transmissibility from $u$ to $v$
such as the viral load and duration of infection of $u$, the
vaccination history and general health of $v$, the duration and nature
of the contact between $u$ and $v$, and characteristics of the
disease.

For each node $u$ we denote its ability to infect others by $\I_u$ and
its ability to be infected by $\Sus_u$.  Each edge has a weight
$w_{uv}$.  The parameter $\alpha$ measures disease-specific
quantities.  In most of our calculations we assume these are scalars
and follow~\cite{miller:heterogeneity,valle:mixing}, setting
\begin{equation}
T_{uv} = T(\I_u,\Sus_v,w_{uv})
      = 1- e^{-\alpha \I_u \Sus_v w_{uv} } \, .
\label{eqn:Tuvweighted}
\end{equation}
If all contacts are identical, $w_{uv}$ may be absorbed into $\alpha$
\begin{equation}
T_{uv} = T(\I_u,\Sus_v) 
      = 1-e^{-\alpha\I_u\Sus_v} \, .
\label{eqn:Tuvunweighted}
\end{equation}
Note that $T_{uv}$ is a number assigned to an edge, while $T(\I_u,\Sus_v)$
is a function which states what the transmissibility between two nodes
would be if they shared an edge.

With mild abuse of notation we denote the probability density
functions (pdfs) of $\I$, $\Sus$, and $w$ by $P(\I)$, $P(\Sus)$, and
$P(w)$ respectively.  We assign $\I$ and $\Sus$ independently, but
allow $w$ to be assigned either independently or based on observed
contacts (\emph{i.e.}, by observing contacts in a population we may
create a static network with edge weights assigned based on the
observed contact).  If $w$ is assigned independently, then it is
possible to eliminate edge weights from the analysis by marginalising
over the distribution of weights.  However, if weights are not
independent (for example work or family contacts tend to have
correlated weights) then the details of the distribution and the
correlations are important.

Given the infectiousness $\I_u$ of node $u$, we
follow~\cite{miller:heterogeneity,miller:bounds} and define its
\emph{out-transmissibility}
\begin{equation}
T_{out}(u) = \iint  T(\I_u,\Sus,w)P(\Sus)P(w) \, d\Sus \, dw \, .
\label{eqn:Toweighted}
\end{equation}
This is the marginalised probability that $u$ infects a randomly
chosen neighbour given $\I_u$.
From the definition of $T_{out}$ and the pdf
$P(\I)$ we can calculate the pdf
$Q_{out}(T_{out})$.  We symmetrically define the
\emph{in-transmissibility} $T_{in}$ and its pdf $Q_{in}(T_{in})$.

We denote the average of a quantity by $\ave{\cdot}$.  The average
transmissibility $\ave{T}$ is
\begin{align}
\ave{T} = \iiint T(\I,\Sus, w) P(\I) P(\Sus) P(w) \, d\I \, d\Sus \, dw \, .
\end{align}

\subsubsection{Epidemic percolation networks}
Rather than studying outbreaks as dynamic processes on networks, we
may consider them in the context of Epidemic Percolation Networks
(EPNs)~\cite{kenah:networkbased,kenah:second,miller:bounds}.  The EPN
framework allows us to study epidemics as static objects and is
useful for quickly estimating $\PE$, $\A$, and $\Ro$.  In this section
we summarise properties of EPNs; more details are provided
in~\cite{kenah:networkbased,miller:bounds,miller:heterogeneity}
and~\ref{app:EPN}.

Once the properties of the nodes and edges are assigned, an EPN
$\mathcal{E}$ is created as follows: We place each node of $G$ into
$\mathcal{E}$.  For each edge $\{u,v\}$ in $G$ we place directed edges
$(u,v)$ and $(v,u)$ into $\mathcal{E}$ independently with probability
$T_{uv}$ and $T_{vu}$ respectively.  The nodes infected in an outbreak
correspond exactly to those nodes that may be reached from the index case
following edges of $\mathcal{E}$.  More specifically, the distribution
of out-components of a node $u$ in different EPN realisations matches
the distribution of outbreaks resulting from different epidemic
realisations in the original model with $u$ as the index case.  It may
be shown that the distributions of out- and in-component sizes give us
information about the probability of nodes to start an epidemic or
become infected in an epidemic.  We will see that in a large
population the structure of a single EPN can be used to accurately estimate
$\PE$, $\A$, and $\Ro$.

Once we create an EPN and choose the index case, we define the
\emph{rank} of node $v$ as the length of the shortest directed path
from the index case to $v$.\footnote{We follow~\cite{ludwig} in using
  the term \emph{rank} rather than \emph{generation} which has been
  used elsewhere, but is potentially ambiguous.  The rank is the
  shortest number of infectious contacts between the index case and a
  node.  It is possible that a different path takes less time.  The
  path infection actually follows is the path that is shorter in
  time, rather than number of links.} If no such path exists, $v$ is
never infected.  

Interchanging all arrow directions interchanges $\PE$ and $\A$.
This means that if we can calculate $\PE$, then $\A$ may be calculated
by the same technique, but with the direction of infection reversed.
Because of this, we focus our attention on calculating $\PE$, and
apply the same methodology to calculate $\A$.  An important
consequence is that if $T$ is constant, then $\PE =
\A$~\cite{newman:spread,miller:heterogeneity}.

\subsubsection{The basic reproductive ratio}
We expect that epidemics are possible if and only if the \emph{basic
  reproductive ratio} $\Ro$ is greater than $1$.  That is, if an
average infection causes more than one new case, an epidemic may
occur, but otherwise the outbreak dies out quickly.  However, this use
of $\Ro$ is not consistent with the typical definition: \emph{the
  average number of new infections caused by a single infected
  individual introduced into a fully susceptible population}, which
gives $\Ro=\ave{T}\ave{k}$.  A more appropriate definition is
\emph{the average number of new infections caused by infected
  individuals early in outbreaks}.  The distinction is subtle, but
results from the fact that whether an outbreak can grow depends on
whether the people of low rank infect more than one person
each~\cite{diekmann}.  Low rank individuals may be different from the
average individual.  Most obviously, they have more
contacts~\cite{newman:spread,feld:friends}; but with clustering, they
also have a disproportionately large fraction of neighbours infected or
recovered.

In order to quantify $\Ro$ more rigorously, we first define $N_r$ to
be the number of people of rank $r$ for a given outbreak simulation.
We then define the \emph{rank reproductive ratio}
\begin{equation}
\Ro[r] = \frac{\E[N_{r+1}]}{\E[N_r]}
\label{eqn:Rog}
\end{equation}
to be the expected number of new cases caused by a rank $r$ node
(averaged over all possible outbreak realisations).  $\Ro[0] =
\ave{T}\ave{k}$ corresponds to the usual definition of $\Ro$.  In
practise, we find that $\Ro[r]$ reaches a plateau quickly as $r$
increases before eventually decreasing as the finite size of the
population becomes important.  Consequently, an improved
definition of $\Ro$ is the limit of $\Ro[r]$ as $r$ grows, subject to
the assumption that $\Ro[r]$ is unaffected by
the finite size of $G$.  This gives (\emph{cf}, \cite{trapman})
\begin{equation}
\Ro=\lim_{r\to\infty} \lim_{|G| \to \infty} \Ro[r] \, .
\label{eqn:Ro}
\end{equation}
and generalises the definition given by~\cite{diekmann} for ODE
models.  Under this definition, epidemics are possible if $\Ro>1$, but
not if $\Ro<1$.  We discuss this further in~\ref{app:R0}.  In a large
population considering multiple index cases with a single EPN gives a
good estimate of $\E[N_r]$ and hence $\Ro[r]$.

\subsection{Configuration Model Networks}

We consider two different types of networks.  The first is a class of
(unclustered) random networks for which we can derive analytic results
based only on the degree distribution.  These analytic results will
form the leading order term of our perturbation expansions.  The
second is a more complicated network resulting from an agent-based
simulation, which we will use to demonstrate the accuracy of our
perturbation expansions.

Our random networks are created by an algorithm which has been
discovered independently a number of times (see \emph{e.g.},
\cite{molloyreed} and~\cite{bollobas:CM}).  These have come to be
called Configuration Model (CM)~\cite{newman:structurereview}
networks.  These networks are maximally random given the degree
distribution.  As the number of nodes in
a CM network grows, the frequency of short cycles becomes negligible.
The resulting lack of clustering allows us to calculate analytic
results for epidemics.  We briefly discuss these results assuming $T$
is constant.  More details are
in~\cite{andersson:limit,newman:spread,meyers:directed,kenah:second,miller:heterogeneity,babak:finite,marder}
and~\ref{app:unclustered_review} (which also addresses edge weights).

In the early stages of an outbreak in a CM network, the probability
that a newly infected (non-index case) node has degree $k$ is
$kP(k)/\ave{k}$.  Clustering is unimportant and so the node will have
$k-1$ susceptible neighbours, regardless of its rank.  Thus the
expected number of infections caused by a newly infected node is
\begin{equation}
\Ro = T\frac{\ave{k^2-k}}{\ave{k}} \, . \label{eqn:unclusteredR0}
\end{equation}

To calculate the probability $\PE$ that infection of a randomly chosen
index case results in an epidemic, we instead calculate the
probability $f=1-\PE$ that it does not.  Then $f$ is the probability
that each neighbour of the index case either is not
infected, or is infected but does not start an epidemic.
Defining $h$ to be the probability that a secondary case
does not start an epidemic, 
\begin{equation}
f = \sum_k P(k) [1-T + Th]^k \, .
\label{eqn:f}
\end{equation}
We find a similar relation for $h$, except that the probability for a
secondary case to have degree $k$ is $kP(k)/\ave{k}$ and only $k-1$
neighbours are susceptible
\begin{equation}
h = \frac{1}{\ave{k}}\sum_k kP(k)[1-T+Th]^{k-1} \, .
\label{eqn:h}
\end{equation}
We solve this recurrence relation for $h$ numerically, and use the
result to find $f$.  $\PE$ follows immediately.  Because $T$ is
constant, this also gives
$\A$~\cite{newman:spread,miller:heterogeneity}.

If $T$ is not constant, the calculation becomes more difficult, and is
discussed further in~\ref{app:unclustered_review}
and~\cite{kenah:second,miller:heterogeneity}.  In general, if $T$ can
vary for CM networks, $\Ro = \ave{T}\ave{k^2-k}/\ave{k}$, while $\PE$
and $\A$ are overestimated by the values calculated assuming
constant $T$.

\subsection{The EpiSimS Network}
We are interested in understanding the impact of clustering on disease
spread.  The term \emph{clustering} is rather vague, and is usually
measured by the number of triangles in a
network~\cite{watts:collective}.  However, any sufficiently short
cycles impact the spread of an infectious disease.  For our purposes
we think of a clustered network as a network with enough short cycles
to impact disease dynamics.

It is relatively simple to measure the degree distribution of a
population using survey methods.  We can easily calculate $\PE$,
$\A$, and $\Ro$ for a CM network with the same degree distribution,
but the error between these values and the values for the original
clustered network are unknown.  Our goal in this paper is to develop
analytical techniques to quantify these errors.

To test our predictions we turn to an agent-based network
derived from a single EpiSimS~\cite{valle:episims,eubanks:episims,barrett}
simulation of Portland, Oregon.  The simulation includes roads,
buildings, and a statistically accurate (based on Census data)
population of approximately 1.6 million people who perform daily tasks
based on population surveys.  This gives a highly detailed knowledge
of the interactions in the synthetic population.  The degree
distribution and contact structure emerge from the simulation.  The
resulting network has significant clustering and average degree of
about 16.  More details are in~\ref{app:episims}.

\section{Clustered networks with homogeneous nodes}
\label{sec:homogeneous}
In this section we assume that the population is homogeneous and all
contacts are equally weighted.  Consequently transmissibility is
constant: $T_{uv}=T$ for all edges.  It follows that $\PE =
\A$~\cite{newman:spread,miller:heterogeneity}.  We develop a predictive theory for $\PE$, $\A$, and $\Ro$ and test the theory with simulations
on the EpiSimS network.  We begin with $\Ro$.

\subsection{The basic reproductive ratio}
\begin{figure}[h]
\begin{center}
\includegraphics[width=0.48\columnwidth]{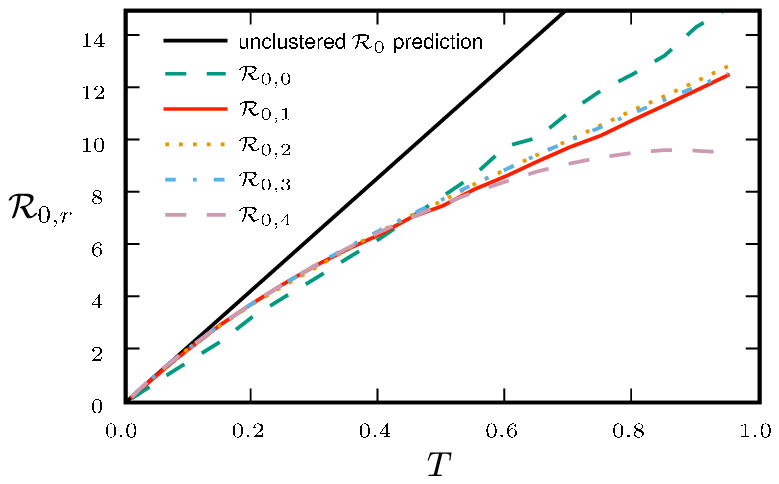}
\hfill\includegraphics[width=0.48\columnwidth]{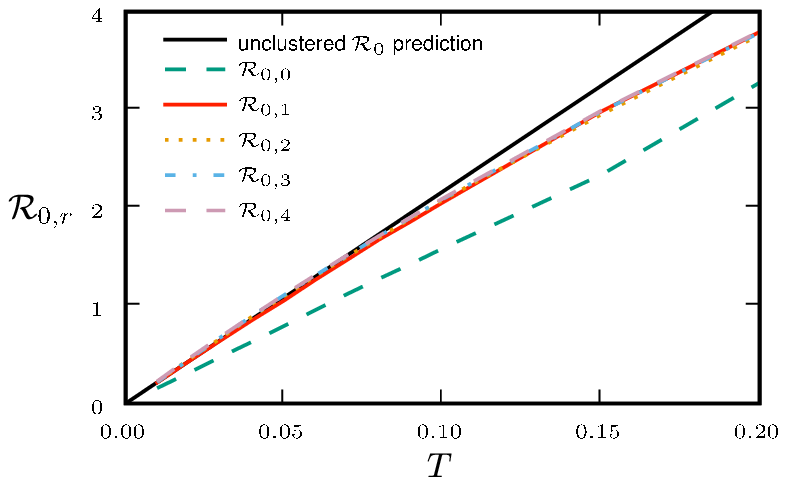}
\end{center}
\caption{Simulated values of the rank reproductive ratio $\Ro[r] =
  \E[N_{r+1}]/\E[N_r]$ for $r=0$, \ldots, $4$ using an EPN from the
  (fixed) EpiSimS network with a homogeneous population, compared with
  the unclustered prediction.  At small $T$ (right panel)
  $\Ro[1]$--$\Ro[4]$ match the unclustered prediction.}
\label{fig:R0g}
\end{figure}

The simulated rank reproductive ratio $\Ro[r]$ is shown in
figure~\ref{fig:R0g} for $0 \leq r \leq 4$.  At all values of $T$,
$\Ro[0] = T\ave{k}$ is clearly distinct from $\Ro[r]$, $r>0$ (which
are close together).  For $r>0$, $\Ro[r]$ is asymptotic to the
unclustered approximation $T\ave{k^2-k}/\ave{k}$ as $T \to 0$.  This
is because at small $T$ the disease only rarely follows all edges of
short cycles and so clustering has no impact.  As $T$ increases, these
curves lie significantly below the unclustered approximation, because
clustering reduces the number of available susceptibles.  $\Ro[4]$
peels away from $\Ro[1]$, $\Ro[2]$, and $\Ro[3]$ for larger $T$
because the population is finite, and so the number of susceptibles
available to infect after rank four is reduced.  In larger
populations, $\Ro[4]$ would not deviate.

\begin{figure}[h]
\begin{center}
\setlength{\unitlength}{0.5pt}
{\begin{picture}(160,160)(0,0)
\put(0,10){$u$}
\put(20,20){\circle*{10}}
\put(20,20){\line(0,1){120}}

\put(20,140){\circle*{10}}
\put(20,140){\line(1,0){120}}

\put(150,140){$v$}
\put(140,140){\circle*{10}}
\put(140,140){\line(0,-1){120}}

\put(140,20){\circle*{10}}
\put(140,20){\line(-1,0){120}}

\put(100,60){\circle*{10}}
\put(20,20){\line(2,1){80}}
\put(100,60){\line(1,2){40}}

\put(60,100){\circle*{10}}
\put(20,20){\line(1,2){40}}
\put(60,100){\line(2,1){80}}

\put(20,20){\line(1,1){120}}

\put(-10,-20){$n_{uv}=4$, \ $\chi_{uv}=1$}
\end{picture}}
\hspace{0.3\columnwidth}
{\begin{picture}(160,160)(0,0)
\put(0,10){$u$}
\put(20,20){\circle*{10}}
\put(20,20){\line(0,1){120}}

\put(20,140){\circle*{10}}
\put(20,140){\line(1,0){120}}

\put(150,140){$v$}
\put(140,140){\circle*{10}}
\put(140,140){\line(0,-1){120}}

\put(100,60){\circle*{10}}
\put(20,20){\line(2,1){80}}
\put(100,60){\line(1,2){40}}

\put(60,100){\circle*{10}}
\put(20,20){\line(1,2){40}}
\put(60,100){\line(2,1){80}}

\put(140,20){\circle*{10}}
\put(140,20){\line(-1,0){120}}
\put(-10,-20){$n_{uv}=4$, \ $\chi_{uv}=0$}
\end{picture}}
\end{center}
\caption{Different options for paths of length two between nodes $u$ and $v$.}
\label{fig:EN2}
\end{figure}
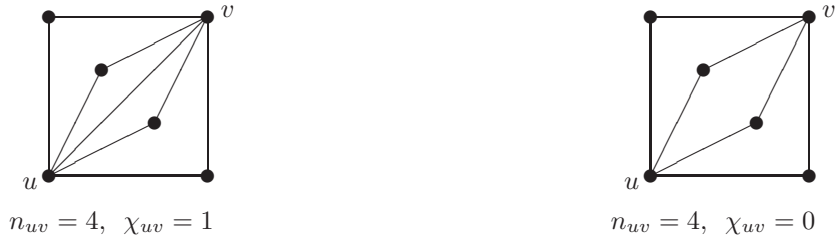

\begin{figure}
\begin{center}
\includegraphics[width=0.48\columnwidth]{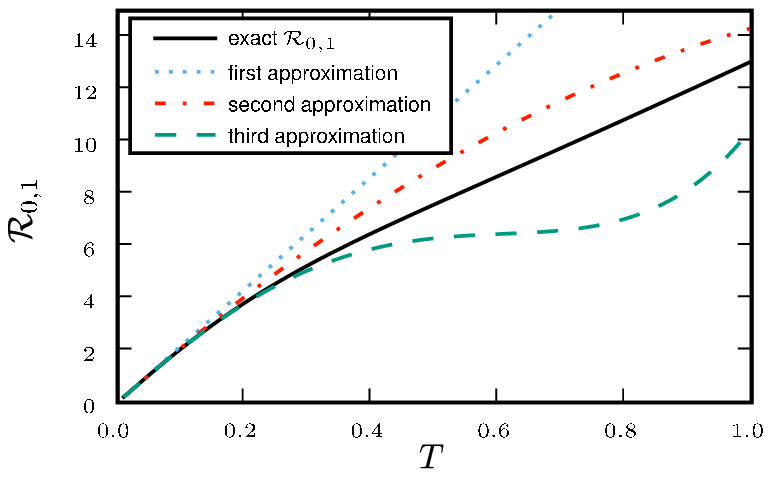}
\hfill
\includegraphics[width=0.48\columnwidth]{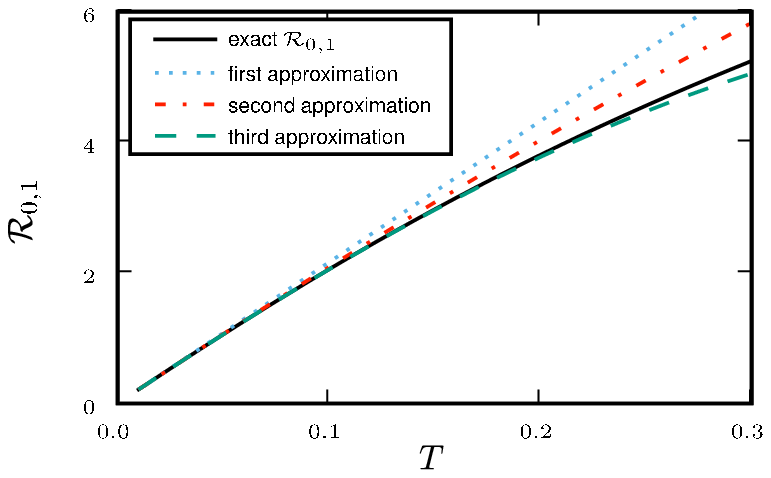}\\
\end{center}
\caption{Comparison of first three asymptotic approximations for
  $\Ro[1]$ from equation~\eqref{eqn:R01homogeneous} with the exact
  value (solid) for the EpiSimS network.  The right panel shows the
  comparison at small $T$.}
\label{fig:R01approx}
\end{figure}

We conclude that $\Ro[r]$ converges quickly, and that $\Ro[1]$ is a
good approximation to $\Ro$, but $\Ro[0]$ is not.  This implies that
the network has important structure contained in paths of length $2$,
but not in paths of length $3$.  This fortunate observation allows us
to approximate $\Ro$ by $\Ro[1]$, which we may analytically calculate
with relative ease ($\Ro[r]$ becomes combinatorially hard as $r$
grows).  To find $\Ro[1]= \E[N_2]/\E[N_1]$ we first note that $\E[N_1]
= T\ave{k}$.  Calculating $\E[N_2]$ is more difficult: consider all
pairs of nodes $u$ and $v$ with at least one path of length $2$
between them.  Let $n_{uv}$ be the number of paths of length $2$
between $u$ and $v$ and $\chi_{uv}$ be an indicator function:
$\chi_{uv}=1$ if $\{u,v\}$ is an edge and $\chi_{uv}=0$ if it is not
(see figure~\ref{fig:EN2}).  The probability that an infection of $u$
results in infection of $v$ in exactly two steps is
$[1-(1-T^2)^{n_{uv}}][1-T]^{\chi_{uv}}$.  Summing this over all pairs
yields
\[
\E[N_2] = \frac{1}{N} \sum_u \sum_{v\neq u} [1-(1-T^2)^{n_{uv}}][1-T]^{\chi_{uv}}
\, ,
\]
(where $N$ is the size of the population and each pair $u$ and $v$
appears twice) which allows us to calculate $\Ro[1]$ exactly.  This
sum is straightforward to calculate, but we can increase our
understanding with a small $T$
expansion.  We approximate $\E[N_2]$ for $T \ll 1$ by
\begin{align*}
  \E[N_2] &=
  \frac{1}{N} \sum_u\sum_{v\neq u} T^2n_{uv}(1-T)^{\chi_{uv}} -
  \binom{n_{uv}}{2} T^4 + \order(T^5) \, ,\\
  &= T^2\ave{k^2-k} - 2T^3\ave{n_\triangle} - T^4\ave{n_\square} +
  \order(T^5) \, ,
\end{align*}
where $\ave{n_\triangle}=\frac{1}{N}\sum_u\sum_{v\neq u} n_{uv}\chi_{uv}$ is the average number of triangles each node is in, and
$\ave{n_\square} = \frac{1}{N} \sum_u \sum_{v\neq u} \binom{n_{uv}}{2}$ is the average number of squares each node is in (\emph{cf},
\cite{hastings:series}).  Higher order terms involve more
complicated shapes.  This gives
\begin{equation}
\Ro[1] = \frac{\ave{k^2-k}}{\ave{k}}T -
\frac{2\ave{n_\triangle}}{\ave{k}}T^2 -
\frac{\ave{n_\square}}{\ave{k}}T^3 + \order\left(\frac{T^4}{\ave{k}}\right) \, .
\label{eqn:R01homogeneous}
\end{equation}
At leading order we recover the unclustered prediction for $\Ro$,
reflecting the fact that at small $T$ the probability 
the outbreak follows all edges of a cycle is negligible.  As $T$
increases, the first corrections are due to triangles, then squares,
then pairs of triangles sharing an edge, and sequentially larger and
larger structures made up of paths of length two.  A comparison of
these approximations with the exact value is shown in
figure~\ref{fig:R01approx}.

\begin{figure}[h]
\begin{center}
 \includegraphics[width=0.48\columnwidth]{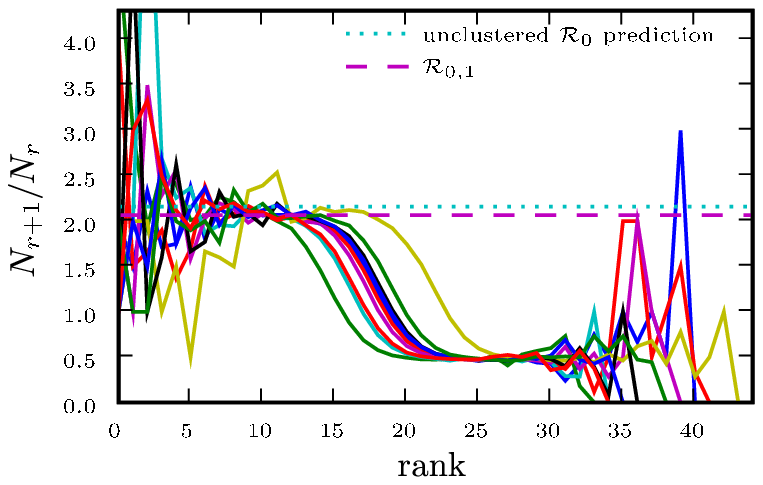}
\hfill
\includegraphics[width=0.48\columnwidth]{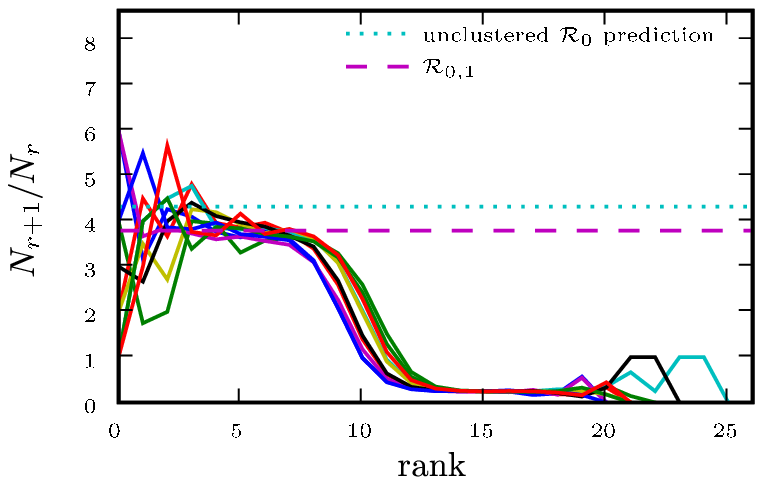}
\hfill 
\end{center}
\caption{The progression of ten simulated epidemics for (left) $T=0.1$
  and (right) $T=0.2$ in the EpiSimS network.  The left panels show
  $N_{r+1}/N_r$ against rank and right panels show the cumulative
  fraction of the population infected.}
\label{fig:simulations}
\end{figure}

Although we have defined $\Ro$ for an ensemble of realisations,
figure~\ref{fig:simulations} shows that $\Ro[1]$ accurately predicts
the observed ratio $N_{r+1}/N_r$ for individual simulations once the
outbreaks are well-established.  Early in outbreaks, the behaviour is
dominated by stochastic effects, and so the ratio of successive rank
sizes is noisy.  Once the outbreak has grown large enough, random
events become unimportant and the ratio settles at $\Ro[1]$.
\footnote{Early noise controls how quickly outbreaks become epidemics,
  and so once stochastic effects become small, the curves appear to be 
  translations in time.  We note that it is common to consider the temporal
  average of a number of outbreaks.  However, prior to taking an
  average, the curves should be shifted in time so that they coincide
  once the stochastic effects are no longer important.  Failure to do
  so underestimates the early growth, peak incidence, and late decay
  while it overestimates the epidemic duration.  This can lead to an
  incorrect understanding of ``typical'' outbreaks.}

\subsection{Epidemic probability and size}
\begin{figure}[h]
\begin{center}
\includegraphics[width=0.8\columnwidth]{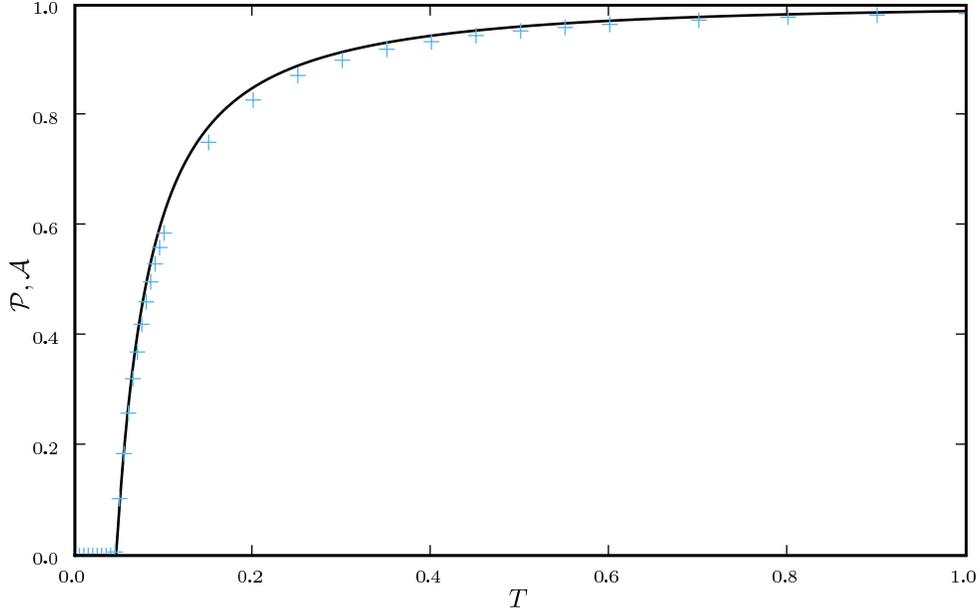}
\end{center}
\caption{Probability $\PE$ and attack rate $\A$ of epidemics for the (clustered)
  EpiSimS network ($+$) versus $T$, compared to the prediction derived
  from the degree distribution assuming no clustering.  Each data
  point is from a single EPN, (the variation in $\PE$ resulting
  from different EPNs is negligible).}
\label{fig:probcomp}
\end{figure}

In order to assess the effect of clustering on $\PE$ and $\A$, we
compare epidemics on the EpiSimS network with the analytic predictions
derived assuming a CM network of the same degree distribution in
figure~\ref{fig:probcomp}.  The epidemic threshold is not
noticeably altered, and the values of $\PE$ and $\A$ are almost
indistinguishable from the predictions made assuming no clustering, despite the large amount of clustering in the network.

Although initially surprising, these results may be understood
intuitively as follows: if $T$ is large enough that the disease
follows all edges of a short cycle then some other edge from a node of
that cycle is likely to start an epidemic and the cycle does not
prevent an epidemic.  On the other hand, if $T$ is smaller so that it
does not follow all edges of a cycle, then the disease never sees the
existence of the cycle, and the outbreak progresses as if there were
no cycle.

To make this more rigorous, we first look at the epidemic threshold.
We assume $\Ro$ is well-approximated by $\Ro[1]$.  Let $T_0=\ave{k}/\ave{k^2-k}$ be the threshold without clustering and $T_0+\delta T$ be the
threshold found by including the correction due to triangles.  From
equation~\eqref{eqn:R01homogeneous} it follows that
\begin{equation}
\frac{\delta T}{T_0} = \frac{2\ave{n_\triangle}\ave{k}}{\ave{k^2-k}^2}
+ \order \left(  \left[\frac{2\ave{n_\triangle}\ave{k}}{\ave{k^2-k}^2}
    \right]^2 \right)\, .
\label{eqn:homog_cond}
\end{equation}
Because a given node of degree $k$ is contained in at most $(k^2-k)/2$
triangles, we conclude $2\ave{n_\triangle}/\ave{k^2-k}\leq 1$.  So if
$\ave{k}/\ave{k^2-k}$ is small the leading order term of
equation~\eqref{eqn:homog_cond} is small and triangles do not
significantly alter the epidemic threshold regardless of the density
of triangles.  For the EpiSimS network,
$\ave{k}/\ave{k^2-k}$ takes the value $0.046$, and 
so we do not anticipate clustering to play an important role in
determining the threshold.

Above threshold, we assume that $\PE$ may be
expanded much like~\eqref{eqn:R01homogeneous}
\begin{equation}
\PE = \PE_0 + \PE_1 \ave{n_\triangle} + \PE_2 \ave{n_\triangle}^2 +\cdots + Q_1 \ave{n_\square} + \cdots \, .
\label{eqn:PEexpansion}
\end{equation}
where $\PE_0$ is the epidemic probability in a CM network of the same
degree distribution.  Although calculating $\Ro[1]$ only requires
information about nodes of distance at most two from the index case,
$\PE$ may depend on effects occurring at
larger distance, and so the expansion has many additional terms.
In general, we expect that if the average degree is large, then the
various coefficients of the correction terms are all small.  The
larger a structure is, the smaller we expect its corresponding
coefficient to be.  The coefficient for triangles $\PE_1$ may
be found by
\[
\PE_1 \ave{n_\triangle}= -\frac{1}{N} \sum_{u \in G} \sum_{\triangle \in G} \pe_\triangle(u) \, ,
\]
where $\pe_\triangle(u)$ is the probability that a given triangle
prevents an epidemic if $u$ is the index case (regardless of whether
$u$ is part of the triangle).  Reversing the order of summation we get
\[
\PE_1 \ave{n_\triangle}= 
-\frac{N_\triangle}{N} \ave{\sum_{u \in G}
     \pe_\triangle(u)}_\triangle = -\frac{1}{3}\ave{n_\triangle} \ave{\sum_{u \in G} \pe_\triangle(u)}_\triangle \, ,
\]
where $N_\triangle$ is the number of triangles in $G$ and
$\ave{\cdot}_\triangle$ is the average of the given quantity taken over
all triangles.  Thus
\[
\PE_1 = - \frac{1}{3}\ave{\sum_{u \in G} \pe_\triangle(u)}_\triangle \, ,
\]
and we can find $\PE_1$ by considering the average effect of a
single triangle in an unclustered network.

To calculate the impact of a triangle with nodes $u$, $v$, and $w$ on
$\PE$ for a given network, we consider that triangle and a randomly
chosen edge $\{x,y\}$ elsewhere in the network.  If we replace the edges
$\{v,w\}$ and $\{x,y\}$ with $\{v,x\}$ and $\{w,y\}$, then
we have a new network without the triangle, but with the same degree
distribution.  We must estimate the expected change in $\PE$ caused by switching the edges.

We begin by assuming $u$ is the index case.  The triangle can affect
$\PE$ only if the infection tries to cross
all three edges, that is, if the infection process `loses' an edge
because of clustering.  This may happen in three distinct ways.  In
the first, node $u$ infects both $v$ and $w$, and then $v$ and/or $w$
tries to infect the other.  In the second $u$ infects $v$ but not $w$,
then $v$ infects $w$, and finally $w$ tries to infect $u$.  The third
is symmetric to the second (with $u$ infecting $w$).

To leading order we can ignore other short cycles, so the probability
that an edge leading out of $u$ (not to $v$ or $w$) will not cause an
epidemic is $g = 1-T + Th$, where $h$ (as before) is the probability
that a randomly chosen secondary case does not cause an epidemic in an
unclustered network and can be calculated using
equation~\eqref{eqn:h}.

\begin{figure}
  \begin{center}
\includegraphics[width=0.4\columnwidth]{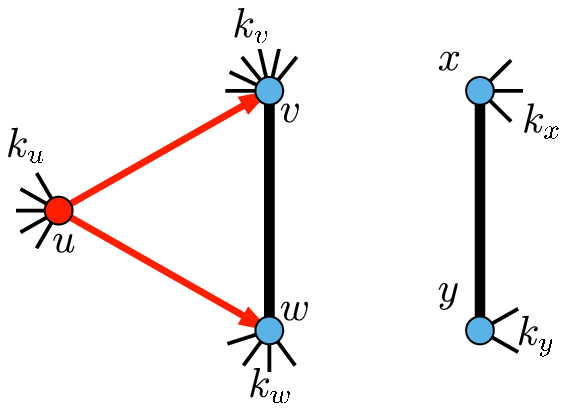}%
\hfill
\includegraphics[width=0.4\columnwidth]{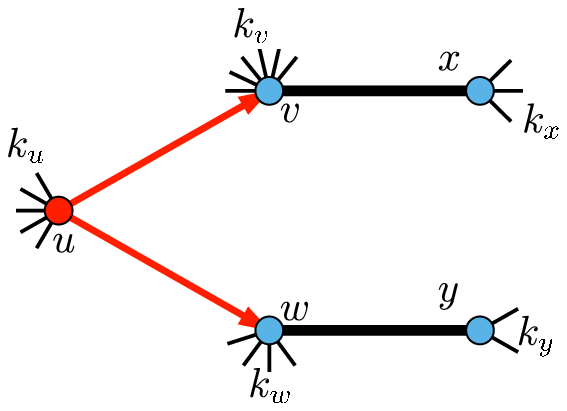}%
\end{center}
\caption{Replacing the edges $\{v,w\}$ and $\{x,y\}$ with $\{v,x\}$
  and $\{w,y\}$ breaks the triangle and allows more infections,
  without affecting the degree distribution.}
  \label{fig:trianglebreak}
\end{figure}

We perform a sample calculation with the first case: $u$ infects both
$v$ and $w$.  Assume that $u$ has degree $k_u$, $v$ has degree $k_v$,
and $w$ has degree $k_w$.  The probability that $u$ infects both $v$
and $w$ without some other edge leading from $u$, $v$, or $w$ starting
an epidemic is $T^2g^{k_u+k_v+k_w-6}$.  If the $\{v,w\}$ edge were
broken and $v$ and $w$ were joined to $x$ and $y$ respectively (see
figure~\ref{fig:trianglebreak}), then the new probability of $u$ to
infect both $v$ and $w$ without an epidemic becomes $T^2
g^{k_u+k_v+k_w-4}$.  The difference is $T^2
g^{k_u+k_v+k_w-6}(1-g^2)$, which is the product of three terms, all at
most $1$.  If the sum $k_u+k_v+k_w$ is moderately large, then
either $g^{k_u+k_v+k_w-6} \ll 1$ or $1-g^2 \ll 1$ (if $g$ is not close
to $1$ then the first term is small, otherwise the second term is
small).  Thus the triangle has little impact on the epidemic
probability in this case.\footnote{If $\PE$
  is small, then the \emph{relative} change may be large, but the
  absolute change is small.}  Similar analysis applies to the
other two cases where the $w$ to $u$ or $v$ to $u$ infections are
lost.  Provided the typical sum of degrees of nodes in a triangle is
relatively large, the probability of an epidemic when the index case
is in the triangle is not impacted significantly.

If the index case is not part of the triangle, then the above analysis
is modified because we must also consider each node in the path from
the index case to the triangle.  We must first calculate the
probability that infection reaches a node in the triangle while
simultaneously no intermediate node sparks an epidemic, and then we
calculate the probability as above that the triangle prevents an
epidemic.  If the index case is $u_1$ and the path from $u_1$ to the
triangle goes through $u_2$, \ldots, $u_n$ and then reaches $u$, then
the probability that the triangle prevents an epidemic $\pe(u_1)$ is
given by $T^n(g^{-2n+\sum_i k_{u_i}})\pe (u)$.  This falls off very
quickly, and so nodes not in the triangle are unimportant, unless
typical degrees are small.


In contrast, in a network with small average degree and a significant
number of triangles this becomes significant.  This explains
observations of~\cite{serrano:prl,serrano:2} who use
networks with average degree less than $3$ and find that clustering
significantly alters $\A$.

It is tempting to generalise our conclusion and state that if the
average degree is large, clustering has no impact on $\PE$ or $\A$.
However, there are a number of counter-examples: consider a network
made up of isolated cliques with $N_c$ nodes, then in
expansion~\eqref{eqn:PEexpansion} the coefficient for cliques of
$N_c$ nodes will not be small.  Consequently care must be taken when
using such an expansion to ensure that neglected terms resulting from
larger scale structures are in fact negligible.  For social networks,
we generally anticipate this highly segregated situation to be unimportant.

We conclude that for most reasonable networks, clustering is only
important for $\PE$ and $\A$ if the typical degrees of nodes are low
in which case $\Ro$ is small.  A consequence of these results is that
if $\Ro$ is moderately large, then $\PE$ and $\A$ are effectively
unaltered by clustering.  If $\Ro$ is small, however, clustering may
or may not play a role in determining $\PE$ and $\A$, depending on
whether $\Ro$ is small because the degrees are small or because $T$ is
small.

\section{Clustered networks with heterogeneous nodes}
\label{sec:heterogeneous}

When we drop the assumption of constant transmissibility, disease
spread becomes more complicated.  If $\I$ is heterogeneous and $u$
infects a neighbour, then the \emph{a posteriori} expectation for
$T_{out}(u)$ becomes higher: it is likely to infect more neighbours.
This accentuates the effect of short cycles, enhancing the impact of
clustering on $\Ro$, $\PE$, and $\A$.  A similar argument applies with
heterogeneity in $\Sus$: if $v$ is not infected by one of its
neighbours, then the \emph{a posteriori} expectation for $T_{in}(v)$
becomes lower: it is less likely to be infected by other neighbours,
and so has multiple opportunities to prevent an epidemic.  Again this
accentuates the effect of short cycles.

In this section we investigate how varying the infectiousness and
susceptibility of nodes in the EpiSimS network enables clustering to
alter the values of $\PE$ and $\A$.  We will make use of the
\emph{ordering assumption} and its consequences
from~\cite{miller:bounds}: if $u_1$ is ``more infectious'' than $u_2$
in a given instance [or $v_1$ ``more susceptible'' than $v_2$], then
$u_1$ is always more infectious than $u_2$ [or $v_1$ always more
susceptible than $v_2$].  More specifically, the ordering assumption
states that if $T_{out}(u_1) > T_{out}(u_2)$, then $T(\I_{u_1},\Sus)
\geq T(\I_{u_2},\Sus)$ for all $\Sus$, and the corresponding statement for $T_{in}$.  The results
of~\cite{miller:bounds} show that if the ordering assumption holds,
heterogeneity tends to reduce $\PE$ and $\A$, and the upper bounds on
$\PE$ and $\A$ correspond to homogeneous populations (constant $T$).

\begin{table}
\begin{center}
\begin{tabular}{c|c|c}
Symbol & Infectiousness & Susceptibility \\
\hline
\di{}&$P(\I)=\delta(\I-1)$ & $P(\Sus)=0.5\delta(\Sus-0.001) + 0.5\delta(\Sus-1)$\\
\sq{}&$P(\I)=0.3\delta(\I-0.001) + 0.7\delta(\I-1)$ & $P(\Sus)=\delta(\Sus-1)$\\
\ex{}&$P(\I)=0.5\delta(\I-0.1)+0.5\delta(\I-1)$ & $P(\Sus)=0.2\delta(\Sus-0.1)+0.8\delta(\Sus-1)$\\
\ci{}&$P(\I)=0.5\delta(\I-0.1)+0.5\delta(\I-1)$ & $P(\Sus)=0.8\delta(\Sus-0.01)+0.2\delta(\Sus-1)$\\
\hline
\pent{}& \text{\parbox{2.2in}{Maximally heterogeneous $P(T_{out}) = \ave{T}\delta(T_{out}-1) + (1-\ave{T})\delta(T_{out})$}} & \text{Homogeneous $T_{in}=\ave{T}$}
\end{tabular}
\end{center}
\caption{For the calculations of sections~\ref{sec:heterogeneous} and~\ref{sec:weighted} we determine $T_{uv}$ using equations~\eqref{eqn:Tuvunweighted} and~\eqref{eqn:Tuvweighted} with the distributions of $\I$ and $\Sus$ given in the first four rows, or by considering a maximally heterogeneous population for which $\ave{T}$ of the population infects all neighbours and $1-\ave{T}$ infect no neighbours.  The function 
  $\delta$ is the Dirac delta function.}
\label{table:symbols}
\end{table}

For simulations in this section, we consider five different
illustrative cases, which will be denoted throughout by the symbol
given in table~\ref{table:symbols}.  In the first four, we use
equation~\eqref{eqn:Tuvunweighted} so that $T_{uv} = 1- e^{-\alpha
  \I_u \Sus_v}$ with the distribution of $\I$ and $\Sus$ varying for
each.  We vary $\alpha$ to change the average transmissibility.  In
the fifth case the out-transmissibility is maximally heterogeneous: A
fraction $\ave{T}$ of the population infect all neighbours, while the
remaining $1-\ave{T}$ infect no neighbours.

The fifth case gives a lower bound on $\PE$ for a homogeneously
susceptible population~\cite{trapman}.  It is hypothesised to remain a
lower bound on $\PE$ if susceptibility is allowed to
vary~\cite{miller:bounds}.  We could also consider maximal
heterogeneity in susceptibility, but the results for $\PE$ and $\A$
merely correspond to interchanging their values for maximal
heterogeneity in infectiousness, and so we do not need to consider it
explicitly.

\subsection{The basic reproductive ratio}
\begin{figure}
\begin{center}
\includegraphics[width=0.48\columnwidth]{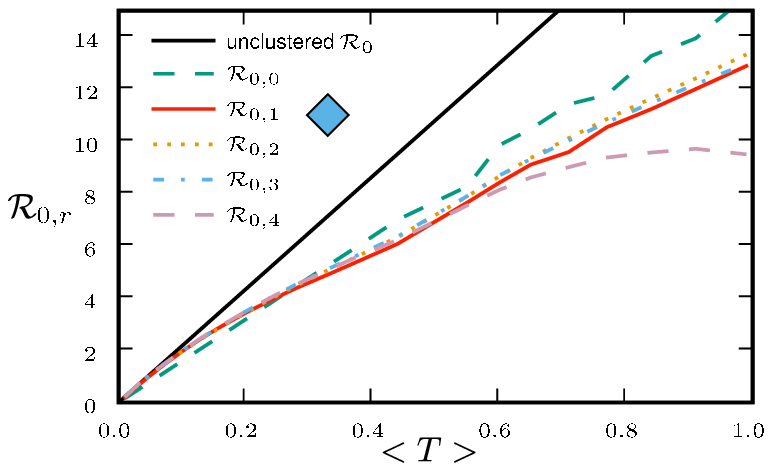}\hfill
\includegraphics[width=0.48\columnwidth]{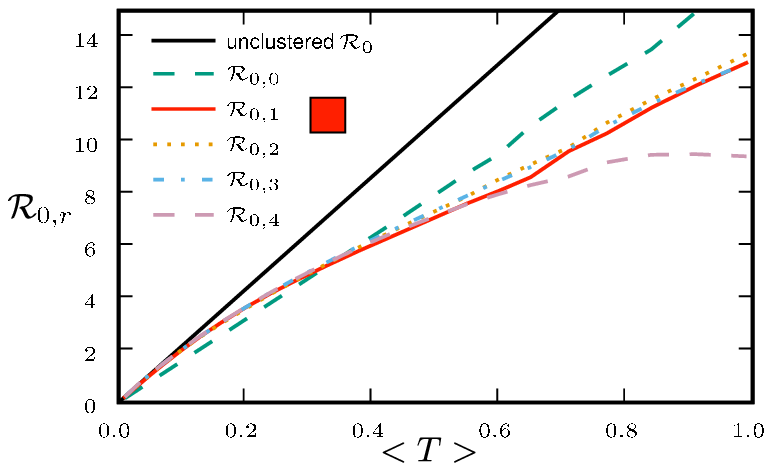}\\\includegraphics[width=0.48\columnwidth]{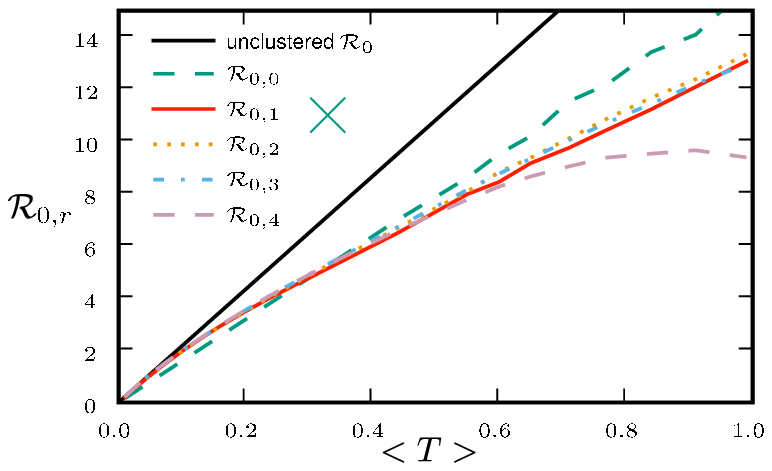}\hfill
\includegraphics[width=0.48\columnwidth]{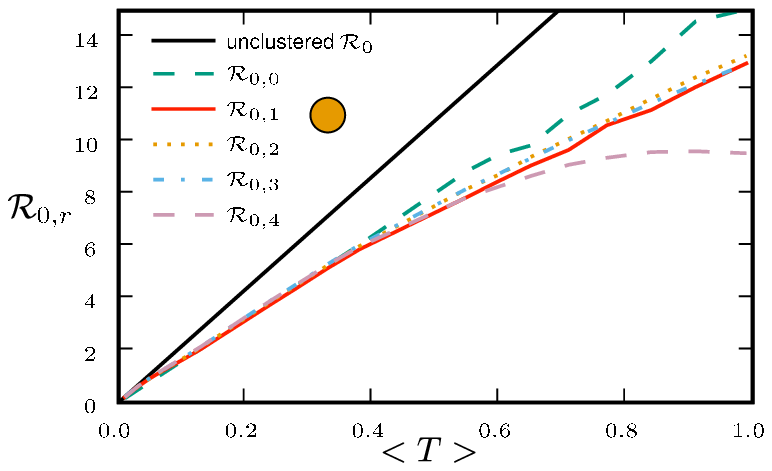}\\
\includegraphics[width = 0.48\columnwidth]{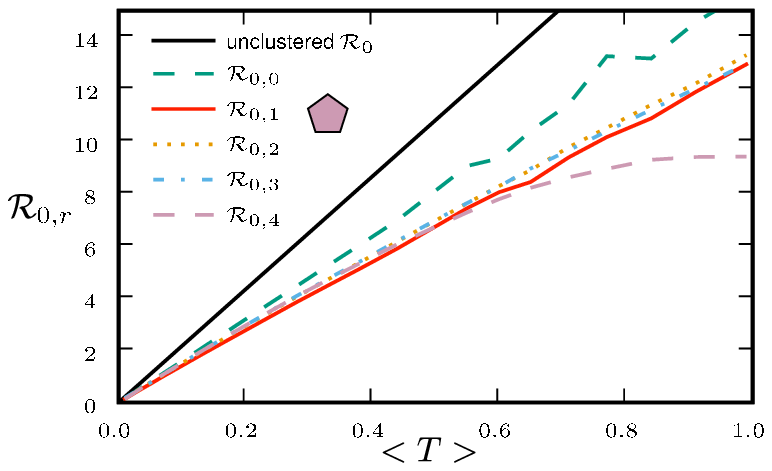}
\hfill
\includegraphics[width = 0.48\columnwidth]{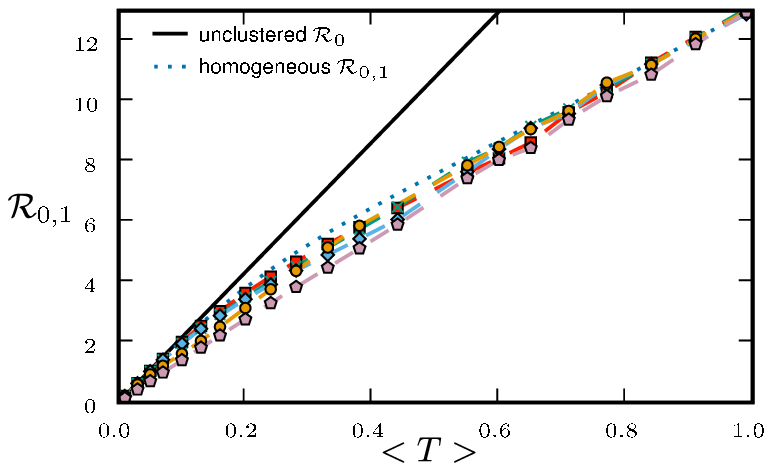}
\end{center}
\caption{$\Ro[r] = \E[N_{r+1}]/\E[N_r]$ calculated from
  simulations for the heterogeneous examples of
  table~\ref{table:symbols}.  The final panel (lower right) compares $\Ro[1]$ for all of the different cases, including both unclustered and homogeneous.}
\label{fig:hetR0g}
\end{figure}

We use simulations to calculate the rank reproductive ratio $\Ro[r]$
for the cases of table~\ref{table:symbols} and plot the result for $0
\leq r \leq 4$ in figure~\ref{fig:hetR0g}.  Note that $\Ro[1]$ remains
a good approximation to $\Ro$.  In the first four cases, 
$\Ro$ is again again asymptotic to the unclustered approximation as
$\ave{T} \to 0$.  There are small kinks for \di{} and \sq{} at
$\ave{T}=0.5$ and $\ave{T}=0.7$ respectively, resulting from the
nature of those distributions.  The heterogeneities act
to enhance the effect of clustering on $\Ro$, but the effect is
relatively small.

In the final, maximally heterogeneous case \pent{}, $\Ro[1]$ remains a good
approximation to $\Ro$.  At small values of $\ave{T}$, the
heterogeneity causes clustering to have a larger impact than in a
homogeneous population as seen in the lower right panel of
figure~\ref{fig:hetR0g}, and so this is not asymptotic to the
unclustered approximation.  At larger values of $\ave{T}$ the
heterogeneous and homogeneous growth rates are similar.

As before, we can calculate $\Ro[1]$ analytically, which helps explain
our observations.  If the ordering assumption holds, we may use a
simplified notation $T(T_{out},T_{in})$ to denote the transmissibility
from a node with out-transmissibility $T_{out}$ to a node with
in-transmissibility $T_{in}$.\footnote{We can use this notation
  because the ordering assumption allows us to uniquely identify $\I$
  from $T_{out}$ and $\Sus$ from $T_{in}$.  If the ordering assumption
  fails, similar results hold, but the notation is more cumbersome.}
We have $\E[N_1] = \ave{T} \ave{k}$ and
\begin{align*}
\E[N_2] &= \frac{1}{N} \sum_u \sum_{v\neq u}
\iint[1-(1-T_{out}T_{in})^{n_{uv}}][1-T(T_{out},T_{in})]^{\chi_{uv}} Q_{out}(T_{out})
Q_{in}(T_{in}) dT_{out} dT_{in}\\
  &=\ave{k^2-k}\ave{T}^2 - 2\ave{n_\triangle} \ave{T_{out}T_{in}T(T_{out},T_{in})}
  - \ave{n_\square} \ave{T_{out}^2}\ave{T_{in}^2} + \cdots \, ,
\end{align*}
and so we may express the growth rate as a perturbation about the
unclustered case $\Ro=\ave{T}\ave{k^2-k}/\ave{k}$ giving
\begin{equation}
\Ro[1] = \frac{\ave{k^2-k}}{\ave{k}} \ave{T} -
\frac{2\ave{n_\triangle}}{\ave{k}}
  \frac{\ave{T_{out}T_{in}T(T_{out},T_{in})}}{\ave{T}} -
  \frac{\ave{n_\square}}{\ave{k}}
  \frac{\ave{T_{out}^2}\ave{T_{in}^2}}{\ave{T}} + \cdots \, .
\label{eqn:R01het}
\end{equation}
For the second term, it may be shown that $\ave{T}^3 \leq \ave{
  T_{out}T_{in}T(T_{out},T_{in})} \leq \ave{T}^2$.  The minimum occurs
when $T$ is constant, suggesting that the maximum growth rate
occurs in a homogeneous population.  The maximum $\ave{T}^2$ occurs
either for \pent{}:
\begin{equation}
Q_{out}(T_{out}) = (1-\ave{T})\delta(T_{out}) +
\ave{T}\delta(T_{out}-1) \, ,
\label{eqn:PoTomax}
\end{equation}
that is, when the out-transmissibility is maximally heterogeneous, or
when the in-transmissibility is maximally heterogeneous:
\begin{equation}
Q_{in}(T_{in})=
(1-\ave{T})\delta(T_{in}) + \ave{T}\delta(T_{in}-1) \, .
\label{eqn:PiTimax}
\end{equation}
Consequently, we expect that for given $\ave{T}$ the minimum growth
rate occurs with maximally heterogeneous infectiousness or
susceptibility.  These two minima for $\Ro[1]$
have previously been hypothesised to give lower bounds on $\PE$ and
$\A$ respectively~\cite{miller:bounds}.

We note that in the maximally heterogeneous case, the correction term
in~\eqref{eqn:R01het} is significant at leading order in $T$.  
Consequently, if $\ave{n_\triangle}$ is comparable to $\ave{k^2-k}/2$
(that is, the clustering coefficient~\cite{watts:collective} is comparable to $1$),
the threshold value of $\ave{T}$ may be increased by clustering, and $\Ro$ is not asymptotic to the unclustered prediction as $\ave{T} \to 0$.

\subsection{Probability and size}
\begin{figure}
 \includegraphics[width = 0.48\columnwidth]{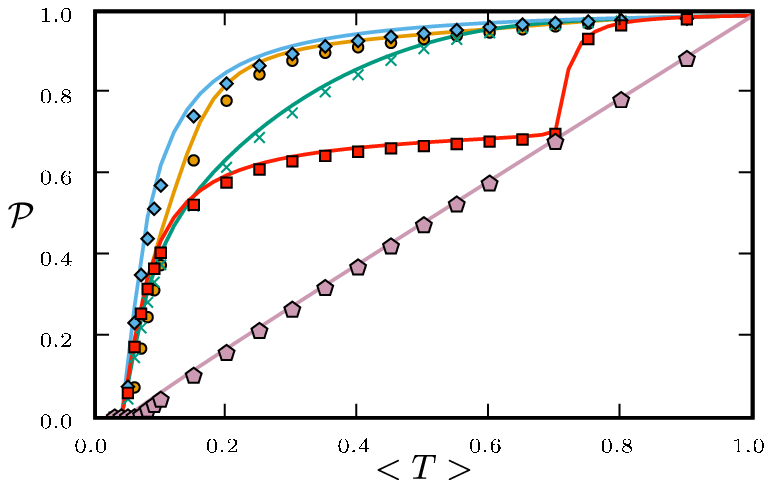} 
 \hfill
 \includegraphics[width = 0.48\columnwidth]{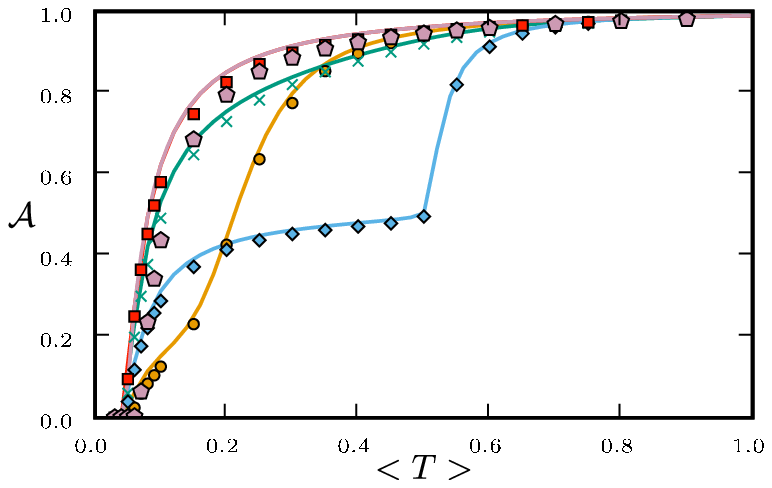}
 \caption{
  Comparison of $\PE$ and $\A$ observed from simulations in the
    clustered EpiSimS network with heterogeneities (symbols) with that
    predicted by the unclustered theory (curves) using
    table~\ref{table:symbols}.  Each data point is based on a single
    EPN.  For both \protect \sq{} and \protect \pent{} $T_{in}(v)=\ave{T}$ for all nodes,
    and so the unclustered prediction for $\A$ is the same.}
 \label{fig:hetsizeprob}
 \end{figure}

Figure~\ref{fig:hetsizeprob} shows that the unclustered predictions
provide a good estimate of $\PE$ and $\A$ in the clustered EpiSimS
network.  We expect that in a network with sufficiently large average
degree, the impact of clustering should once again be small.

We use arguments similar to before, taking a triangle with nodes $u$,
$v$, and $w$.  The reasoning becomes more difficult because knowledge
that $u$ infects $v$ may increase the expectation that $u$ infects
$w$.  Consequently the lost edges in triangles are more frequently
encountered by the outbreak.  However, the knowledge that $u$ infects
$v$ also increases the expectation that $u$ infects its other
neighbours.  For a triangle to prevent an epidemic, we need
both that no edge outside the triangle leads to an epidemic and that
the lost edge would otherwise have caused an epidemic.  If the typical
degree of the network is not small, then the fact that the lost edge
is encountered more frequently may be offset by the fact that when it
is encountered, other edges are more likely to spark an epidemic.

For \pent{} where nodes infect all or none of their neighbours, the
effect of different triangles that share the index case cannot be
separated as easily.  The probability the index case directly infects
a set of $m$ nodes of interest is $\ave{T}$, rather than $T^m$.  Thus
expansions as in~\eqref{eqn:PEexpansion} do not work as well: terms
that were previously higher order become significant.  Close to the
epidemic threshold, this can play an important role.  However, well
above the epidemic threshold, if the index case infects all of its
neighbours, an epidemic is almost guaranteed and so $\PE \approx
\ave{T}$ regardless of whether the network is clustered.  Thus for
\pent{}, clustering affects $\PE$ only close to the epidemic
threshold.

In the opposite case where nodes would be infected by any neighbour or
else no neighbour, the values of $\PE$ and $\A$ are interchanged.
Thus for maximally heterogeneous susceptibility $\PE$ could
be significantly altered close to the threshold.  The reason for this
is as follows: For the first step the spread is indistinguishable from that of an outbreak with constant $T$.  However, when infections
of rank 1 attempt to infect their neighbours, they cannot infect any
of the neighbours of the index case.  In contrast, in the constant $T$
case, any neighbour not infected by the index case would be
susceptible at later steps.  Consequently, the impact of triangles
becomes much more important (by a factor of $1/\ave{T}$) and our
earlier argument for neglecting them fails.  The interaction of
maximal heterogeneity with clustering in this case is larger, but it
nevertheless becomes unimportant far from the threshold.

Our prediction that heterogeneity allows
clustering to be more significant close to the threshold is borne out
for \ci{} where there is relatively strong heterogeneity in
susceptibility just above the epidemic threshold.  The epidemic
threshold for \ci{} is increased compared to the other cases.  In
contrast there is much stronger heterogeneity in susceptibility for
\di{} at $\ave{T}=0.5$ and in infectiousness for \sq{} at
$\ave{T}=0.7$.  This results in a reduction in $\A$ and $\PE$
respectively, but because it is far from threshold, there is little
deviation from the unclustered predictions.

\section{Clustered networks with weighted edges}
\label{sec:weighted}

When we allow edges to be weighted, new complications arise.  The
weights we use in our simulations are the durations of contacts from
the EpiSimS simulation and are discussed in detail
in~\ref{app:episims}.  If a contact in the original EpiSimS simulation
is longer, a higher weight is assigned.  If the weights of different
edges were independent, then we could simply take $T_{uv} = \int
T(\I_u,\Sus_v,w) P(w) \, dw$.  However, edge weights are not
independent: clustered connections tend to have larger weights.  If
brief contacts are negligible, the disease spreads on a subnetwork of
the original network.  The new network has a comparable number of
short cycles to the original, but lower typical degree.  This should
enhance the impact of clustering.

For our calculations in this section, we first isolate the impact of
weighted edges by taking a homogeneous population ($\I=\Sus=1$) and
using $T_{uv}=1-e^{-\alpha w_{uv}}$.  We vary $\alpha$ in order to
set $\ave{T}$.  We then investigate a heterogeneous population using
equation~\eqref{eqn:Tuvweighted} with the first four distributions of
table~\ref{table:symbols}.

\begin{figure}
 \includegraphics[width = 0.48\columnwidth]{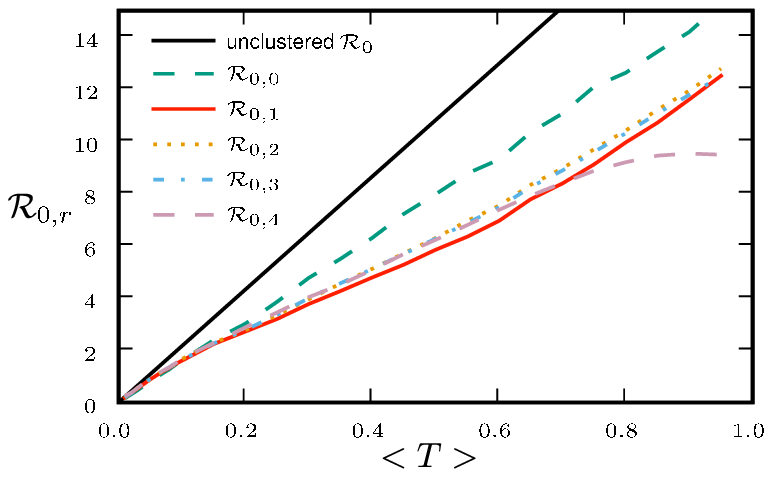}
\hfill
\includegraphics[width = 0.48\columnwidth]{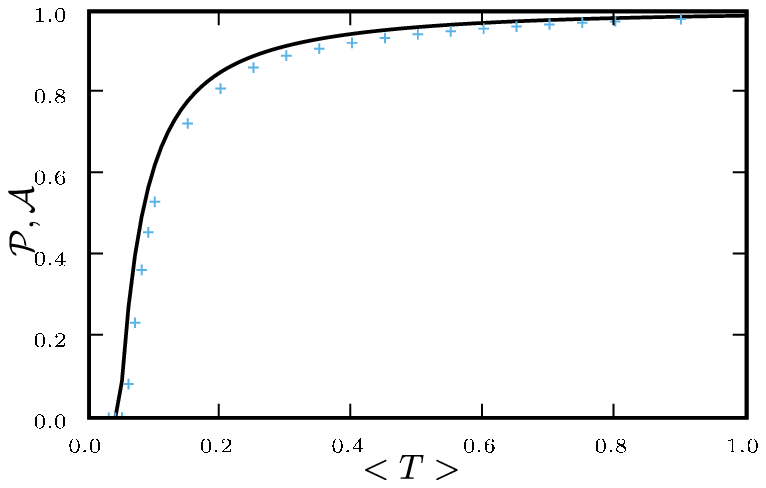}
\caption{$\Ro[r]$, $\PE$, and $\A$ for the weighted EpiSimS network with a
  homogeneous population.}
\label{fig:hom_weighted}
\end{figure}

Results for a homogeneous population are shown in
figure~\ref{fig:hom_weighted}.  Because $T_{uv}=T_{vu}$ for all pairs,
it follows that $\PE = \A$.  If different edge weights were
uncorrelated, then the value of $\Ro$ would match with
figure~\ref{fig:R0g} and $\PE$ and $\A$ would match with
figure~\ref{fig:probcomp}.  We see, however, that $\Ro$ is
significantly reduced from the homogeneous unweighted population (but
$\Ro[1]$ remains a good approximation).  Close to the threshold $\PE$
and $\A$ are mildly reduced.  These observations are consistent with
our expectation that clustering should be accentuated by incorporating
edge weights.  Although the predictions for $\PE$ and $\A$ are not far
off, we expect that they would improve if we adjusted the degree
distribution to match that of the effective network on which the
disease spreads.

\begin{figure}
\begin{center} 
\includegraphics[width=0.48\columnwidth]{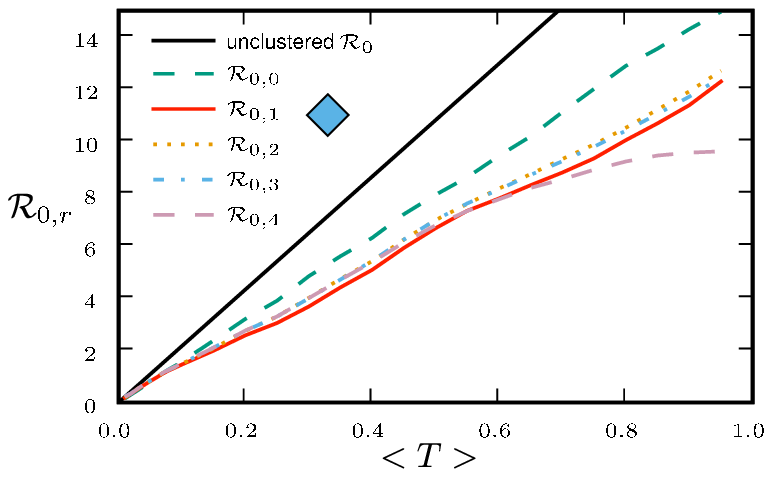}\hfill
\includegraphics[width=0.48\columnwidth]{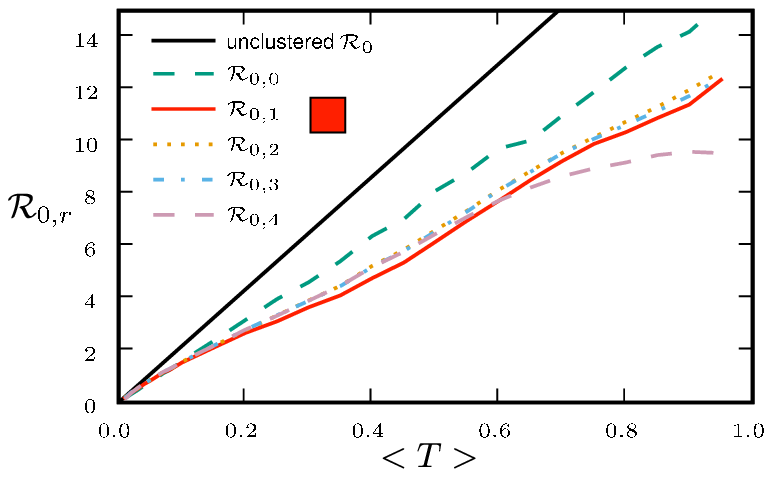}\\
\includegraphics[width=0.48\columnwidth]{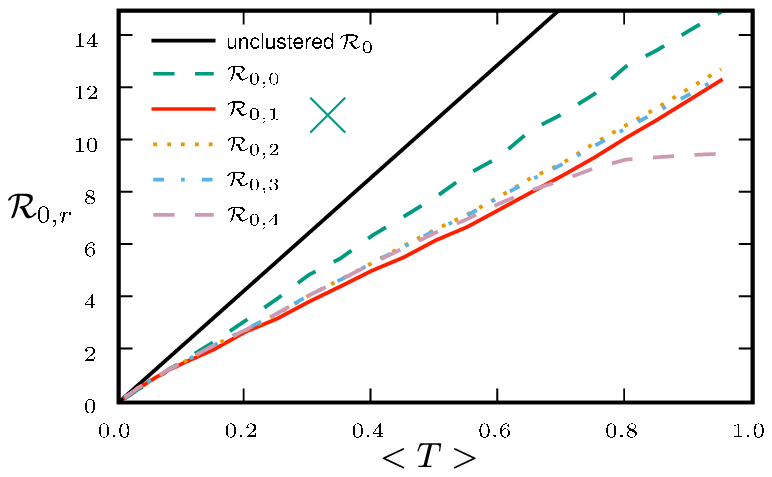}\hfill
\includegraphics[width=0.48\columnwidth]{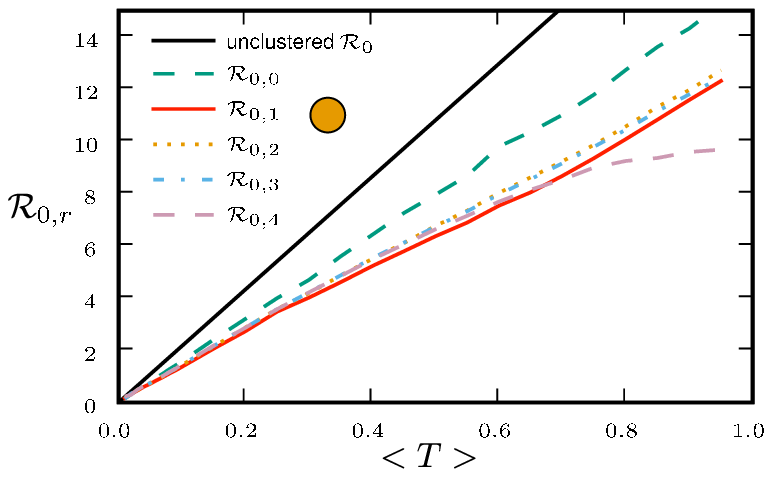}
\end{center}
 \caption{$\Ro[r]$ with heterogeneous transmissibility and weighted
  edges on the EpiSimS network.}
\label{fig:het_weightedR0}
\end{figure}

When the population is moderately heterogeneous
(figure~\ref{fig:het_weightedR0}), we still find that $\Ro[1]$ is a
reasonable approximation to the true value of $\Ro$, however, it
slightly underestimates $\Ro$ as $\ave{T}$ grows.  Unfortunately the analytic
calculation of $\Ro[1]$ is much more difficult, and so it is more
appropriate to use simulations to estimate its value.  If there were
no correlation between weights of different edges, then the
calculation would reduce to that of  section~\ref{sec:heterogeneous}.

\begin{figure}
\includegraphics[width=0.48\columnwidth]{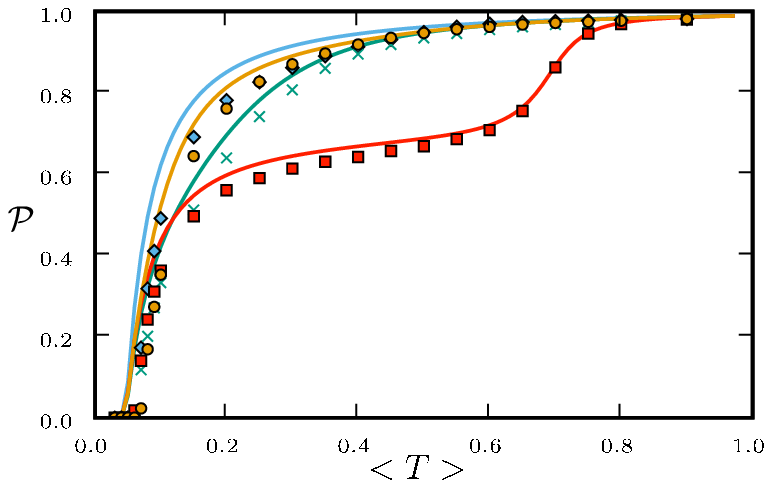} \hfill
\includegraphics[width=0.48\columnwidth]{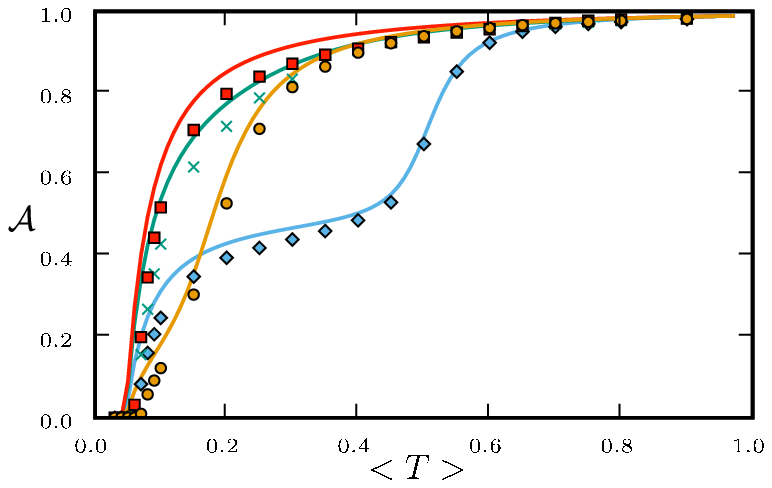}
\caption{Simulated $\PE$ and $\A$ (symbols) for the weighted EpiSims
  network compared with predictions in unclustered networks with
  the same edge weight distribution (curves).}
\label{fig:weightedsizeprob}
\end{figure}

We consider $\PE$ and $\A$ in figure~\ref{fig:weightedsizeprob}.  The
unclustered predictions are reasonable approximations of the actual
values.  The error is larger than before because we have combined two
effects (edge weights and heterogeneity) that both accentuate the
impact of clustering.  In spite of this, the predicted values of $\PE$
and $\A$ are not far off, and the direction of the error is
consistent: the unclustered prediction is always an overestimate.

\section{Discussion}
\label{sec:discussion}

We have investigated the interplay of clustering, node heterogeneity,
and edge weights on the growth rate $\Ro$, probability $\PE$, and size of epidemics $\A$ in
social networks.  For unclustered networks with independently
distributed edge weights, it is possible to predict all these
quantities analytically.  Under weak assumptions we can accurately
estimate $\Ro$, $\PE$, and $\A$ for clustered networks.

If the typical degrees are not small, then for a given average
transmissibility and degree distribution:
\begin{itemize}
\item   The dominant effect
controlling the growth rate of epidemics is clustering.  Increased
clustering reduces $\Ro$.
\item The dominant
effect controlling the probability of epidemics is heterogeneity in
infectiousness.  Increased heterogeneity reduces $\PE$
\item The dominant effect controlling the size of epidemics
is heterogeneity in susceptibility.  Increased heterogeneity reduces
$\A$
\end{itemize}
We are thus able to neglect clustering and still closely estimate
$\PE$ based only on the degree distribution and the
out-transmissibility pdf $Q_{out}$.  The estimate for $\A$ depends
only on degree distribution and the in-transmissibility pdf $Q_{in}$.
The impact of clustering is significant in altering $\Ro$, and its
impact is mildly enhanced by heterogeneities.  This enhancement occurs
because the probability of following all edges of a cycle is increased
if some of the edges are correlated due to the heterogeneity.  If
heterogeneity is large, clustering may play a small role in moving the
epidemic threshold, but otherwise its effect on the threshold is
negligible.  In networks with small typical degree, it has been
observed that clustering can modify $\PE$ or
$\A$~\cite{serrano:prl,serrano:2}, which is consistent with our
estimates.

If edge weights are included, but are independently distributed, then
their impact is in modifying $Q_{in}(T_{in})$ and $Q_{out}(T_{out})$.
The resulting modification may be calculated explicitly, and edge
weights have no further effect.  If edge weights are correlated, they
have a more important role in governing the behaviour of epidemics,
particularly if higher weight edges tend to be the clustered edges (as
frequently occurs in social networks).  If this happens, then the
impact of clustering is enhanced, and the growth rate of epidemics is
further reduced.

When we move from predicting $\PE$ and $\A$ to predicting $\Ro$, we
find that the growth rate is well approximated by $\Ro[1] =
\E[N_2]/\E[N_1]$.  This may be calculated analytically in the
homogeneous case (constant $T$).  When heterogeneities are included,
the calculation becomes harder, and when edge weights are included it
becomes largely intractable.  However, these are easily estimated
through simulation.

These observations show that using $\Ro$ to predict $\A$ will
generally be inadequate.  In a homogeneous but clustered population,
$\Ro$ is reduced but $\A$ is unaffected, and so predictions of $\A$
based on $\Ro$ will be too small.  In networks that are not clustered
but have heterogeneities in susceptibility, $\Ro$ is unaffected but
$\A$ is substantially reduced.  Consequently, the value of $\A$
predicted from $\Ro$ will be too large.

Perhaps our most important conclusion about clustering is that it
plays an important role in altering the growth of an epidemic, but it
only plays a small role in determining whether an epidemic may occur
or how big it would be.  If the relevant questions are, ``how likely is
an epidemic and how large would it be?''\ then the modeller may proceed
ignoring clustering.  If however, the question is ``how fast will an
epidemic grow?''\ then clustering must be considered, but only enough
to calculate $\Ro[1]$.

Our results have implications for designing intervention strategies.
A number of strategies are available to control epidemic spread,
including travel restrictions, quarantines, and vaccination.  Most of
the mathematical theory predicting the effects of these strategies has
been developed under the assumption of no clustering.  Most
immediately, if we measure $\Ro=2$ at the early stages of an epidemic,
traditional approaches will suggest that vaccinating just over half of
the population will bring the epidemic below threshold.  However, if
the population is clustered, then the observed $\Ro$ was already
affected by the fact that some transmission chains were redundant.
Following vaccination, some of these chains will no longer be
redundant and the disease may still spread with $\Ro>1$.

Achieving a better understanding of the effect of clustering further
helps to guide our intuition when choosing between strategies.  For
example, let us assume that we have the choice between two strategies:
in the first, we stagger work schedules in such a way that a typical
person's contacts is reduced by $1/3$; in the second, we implement
population-wide behavior changes so that the same reduction in number
of contacts is achieved, but the work contacts are unaltered.  The
first reduces clustering while the second increases the relative
frequency of clustering.  The value of $\Ro$ is much smaller in the
second case than in the first because of the larger clustering, but
$\PE$ and $\A$ are reduced by a comparable amount in both cases.
Which strategy is best depends on our goals and relative costs.

Strategies that enhance heterogeneity in infectiousness or
susceptibility can be important to help reduce $\PE$ or $\A$, even
when there is little impact on $\Ro$.  Depending on which quantity we
want to minimize, different choices will be optimal.  Consider a
choice between vaccinating all individuals with a vaccine that reduces
$T_{uv}$ by a factor of $1/2$ for all pairs $u$ and $v$ or a contact
tracing strategy that will remove $1/2$ of all new infections before
they have a chance to infect anyone.  Both strategies reduce $\ave{T}$
by a half.  However, the first reduces $T_{out}$ uniformly, while the
second increases heterogeneity in $T_{out}$.  Thus if we have the
choice of the two strategies, contact tracing is more likely to
eliminate the disease before an epidemic can happen.  If our choice is
instead between a global vaccine reducing $T_{in}$ by a factor of
$1/2$ for all individuals, or a completely effective vaccine that is
only available for $1/2$ of the population, the latter choice will be
more effective for reducing $\A$.

\section*{Acknowledgements}
  This work was supported by the Division of Mathematical Modeling at
  the UBC CDC under CIHR (grants no. MOP-81273 and PPR-79231) and the
  BC Ministry of Health (Pandemic Preparedness Modeling Project), by
  DOE at LANL under Contract DE-AC52-06NA25396 and the DOE Office of
  ASCR program in Applied Mathematical Sciences, and by the RAPIDD
  program of the Science \& Technology Directorate, Department of
  Homeland Security and the Fogarty International Center, National
  Institutes of Health.  Lu\'{\i}s M. A. Bettencourt contributed
  greatly to the early development of this work.  I am grateful to
  Sara Y del Valle for providing the EpiSimS network data.

\appendix
\section{Epidemic Percolation Networks}
\label{app:EPN}

\begin{figure}
\begin{center}
 \includegraphics[width=0.48\columnwidth]{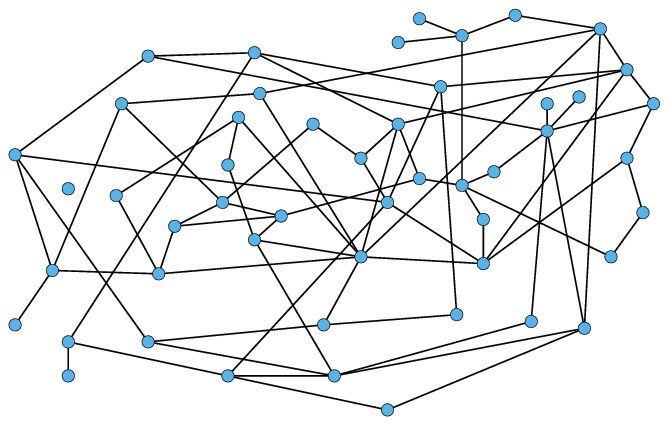}\hfill
\includegraphics[width=0.48\columnwidth]{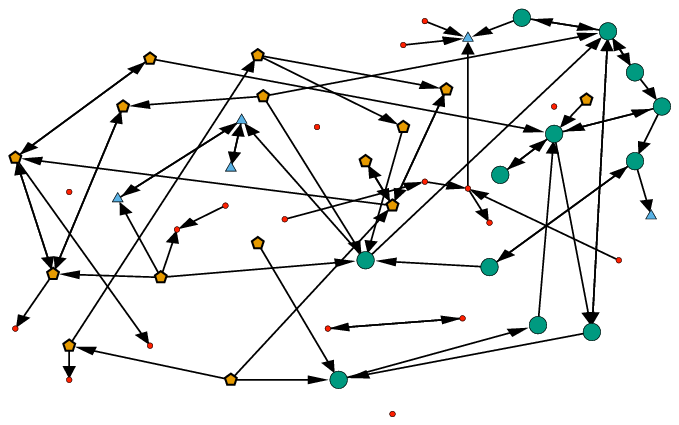}
\caption{The underlying network for figure~\ref{fig:sample}
  and an EPN that leads to the same outbreak.  Nodes in the $G_{scc}$
are denoted by large circles, nodes in the $G_{in}$ (but not in the
$G_{scc}$) are denoted by pentagons, nodes in the $G_{out}$ (but not in
the $G_{scc}$) are denoted by triangles, and nodes not in any of these
components are denoted by small circles.}
\label{fig:epn}
\end{center}
\end{figure}

In this appendix, we describe the \emph{Epidemic Percolation Network}
(EPN), a tool that allows us to consider an epidemic as a static
object rather than a dynamically changing process.  This eases the
understanding of certain key features and provides an improved
technique to efficiently estimate $\PE$.
EPNs have received moderate use
recently~\cite{kenah:second,kenah:networkbased,miller:bounds}, and a
precursor appeared in~\cite{ludwig}.  A sample EPN for an
\erdosrenyi{} network of average degree $3$ and $T=0.4$ is shown in
figure~\ref{fig:epn}.

\begin{figure}[h]
\includegraphics[width=0.48\columnwidth]{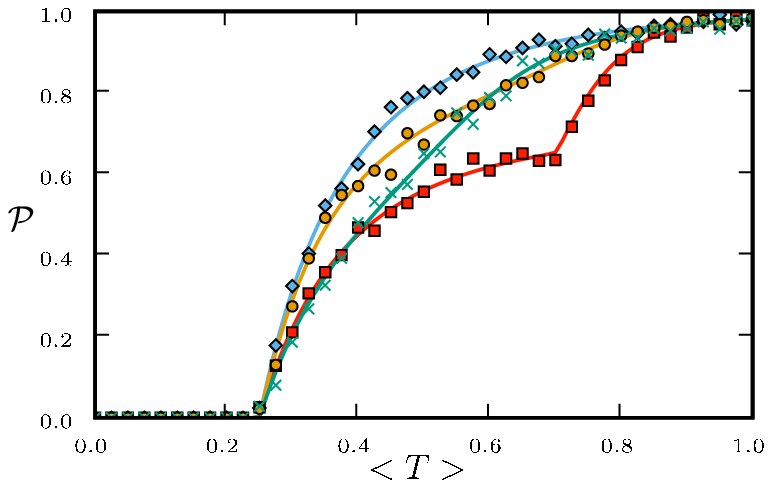}
\hfill
\includegraphics[width=0.48\columnwidth]{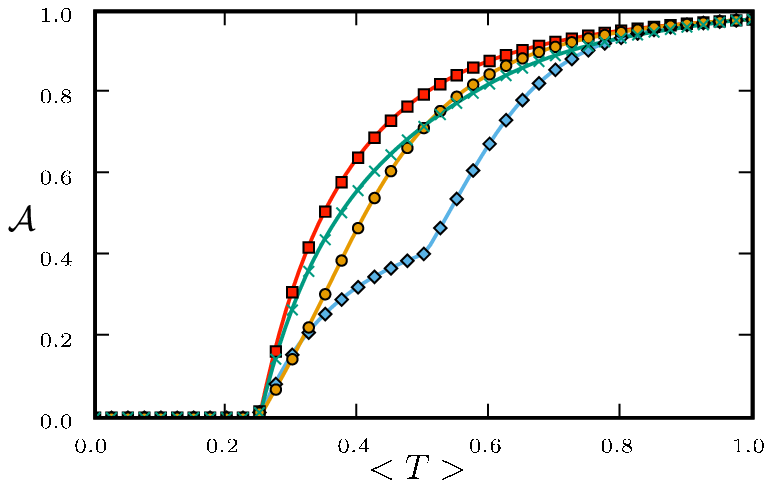}
\caption{$\PE$ and $\A$ in an \erdosrenyi{} network
  of $10^5$ nodes and $\ave{k}=4$.  Theory (curves) compare well with
  results of $500$ simulations (symbols).  We take $T_{uv} =
  1-e^{-\alpha \I_u\Sus_v}$, with distributions of $\I$ and $\Sus$
  as given in table~\ref{table:symbols}.}
\label{fig:er_vb_sim}
\end{figure}

Typically to estimate $\PE$ in an SIR model many Monte Carlo simulations are
performed.  This process requires many iterations to have confidence
in the results.  Representative results from $500$ such simulations
are found in figure~\ref{fig:er_vb_sim}.  Note that there is
considerably more noise in the estimates of $\PE$ than in the
estimates of $\A$.

Instead we generate a single EPN $\mathcal{E}$.  We first assign $\I$
and $\Sus$ to each node and (if necessary) $w$ to each
edge.\footnote{It is important that this assignment occur prior to
  infection [or at least independent of outbreak history].  If the
  infectiousness of $v$ depends on the infectiousness of the node that
  infected $v$, then these results fail.  This is the
  \emph{time-homogeneity} assumption of~\cite{kenah:networkbased} and
  is also used by~\cite{ludwig}.  Some effects that can occur when
  this assumption is false appear in~\cite{floyd:SIR}.}  Then for each
node $u$ and neighbour $v$ we calculate $T_{uv}$ and place the directed
edge $(u,v)$ into $\mathcal{E}$ with probability $T_{uv}$.  The
distribution of out-components of a given node is the same as for the
final outbreak following an introduced infection of that node in the
original epidemic model.

\begin{figure}

  \includegraphics[width=0.48\columnwidth]{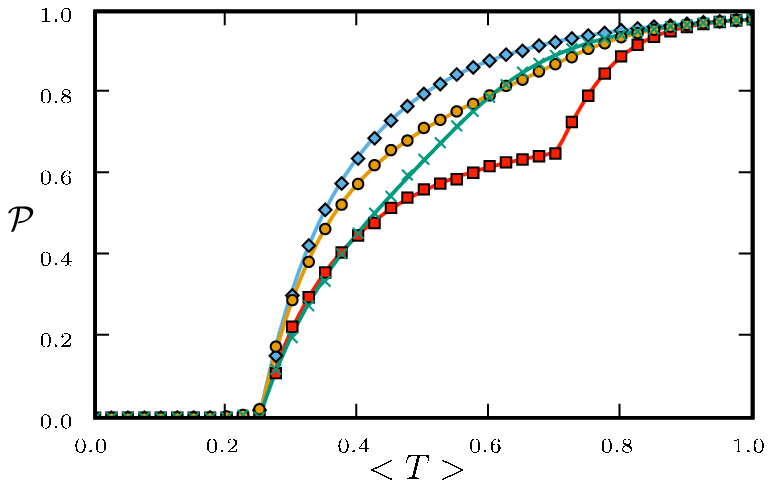}
  \hfill
  \includegraphics[width=0.48\columnwidth]{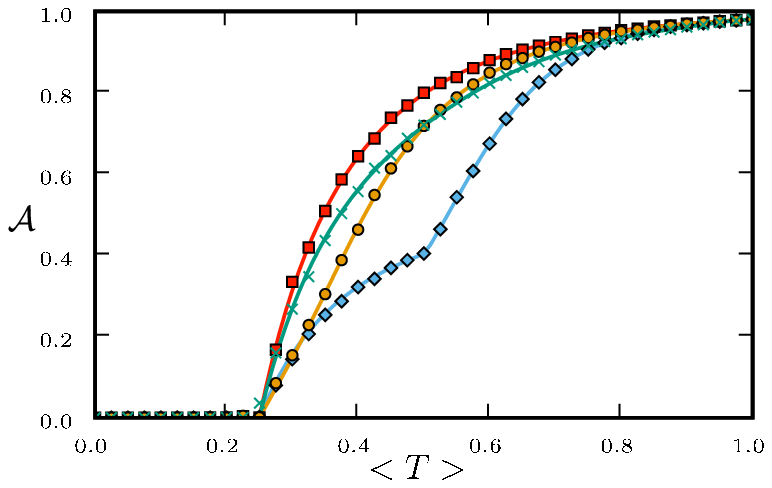}

\caption{Same as figure~\ref{fig:er_vb_sim}, but calculated through a
  single EPN for each $T$.  The noise is substantially reduced in the $\PE$
  calculations, but slightly increased in the $\A$ calculations.}
\label{fig:er_vb_epn}
\end{figure}

If the system is above the epidemic threshold, then $\mathcal{E}$ will
(almost surely) have a giant strongly connected component
$G_{scc}$~\cite{broder,dorogovtsev}.  We follow~\cite{dorogovtsev} and
define the set of nodes (including $G_{scc}$) from which $G_{scc}$ may
be reached following the directed edges to be the giant in-component
$G_{in}$.  We symmetrically define $G_{out}$ to be the set of nodes
reachable from $G_{scc}$.  Note that $G_{scc} = G_{in} \cap G_{out}$.
If the initial infection is in $G_{in}$, an epidemic occurs, and all
nodes in $G_{out}$ become infected.  Thus the size of $G_{in}$
corresponds to the probability of an epidemic $\PE$ and the size of
$G_{out}$ corresponds to the size of an epidemic $\A$.  This may be
seen by comparing the EPN in figure~\ref{fig:epn} with the outbreak
shown in figure~\ref{fig:sample}.\footnote{It is possible that a small number
of nodes outside of $G_{out}$ are infected, but the proportion
vanishes as $|G| \to \infty$.}

Thus in the limit of large networks, epidemic probability is
well-approximated by $\PE=|G_{in}|/|G|$ while the fraction infected is
well-approximated by $\A = |G_{out}|/|G|$.  This observation allows us
to estimate $\PE$ from a single EPN
(figure~\ref{fig:er_vb_epn}), rather than from hundreds of simulations
(figure~\ref{fig:er_vb_sim}).  If the structure of the network is
sufficiently random, the error in $\PE$ and $\A$ from a single EPN is
$\order(\log N/N)$ (see, \emph{e.g.},~\cite{bollobas:random}), and so
in a large population a single simulation will provide a sufficiently
good estimate.

\section{The basic reproductive ratio}
\label{app:R0}
In this appendix we provide examples demonstrating the need for the more careful
definition of $\Ro$ in section~\ref{sec:formulation}, and we explore properties
of this definition.

A pair of simple examples demonstrates the difficulties with the
standard definition.  In our first example, the standard definition
suggests no epidemic is possible ($\Ro<1$), while in fact they are.
In our second example, the standard definition suggests epidemics are
possible ($\Ro>1$), while in fact they are not.

For the first example, consider a fully-connected population of $|G|
\gg 1$ nodes.  We add $3|G|$ isolated nodes and consider a disease for
which $T=3/|G|$ .  A node in the connected component will infect on
average $3$ nodes, while an isolated node infects none.  On average
therefore, a random index case infects $0.75$ other nodes.  Under the
standard definition $\Ro=0.75$ and epidemics should be impossible.
However, if the index case is in the connected component, the
introduction is likely to lead to an epidemic.

Alternately, consider a population of $|G|$
nodes with each node having three neighbours.  For simplicity we assume
no short cycles.  Assume that a disease spreads with probability $p
\in (1/3,1/2)$ to a given neighbour.  The average number of secondary
infections caused by a single introduced infection is $3p>1$, giving
$\Ro>1$ under the standard definition.  However, each secondary
infection has only two susceptible neighbours, and so infects on
average $2p<1$ neighbours, and the outbreak dies out.

Some of these issues have been dealt with by~\cite{diekmann}, who
considered compartmental deterministic models of several types of
individuals.  At early time nonlinear terms are unimportant, and the
profile of the infected population aligns with the eigenvector of the ``next-generation'' matrix.  In stochastic settings, the same alignment occurs, but
it may do so more quickly or slowly than predicted and for some
realisations it may instead die out.  To make a more rigorous
definition of $\Ro$, we turn to statements about the average behaviour.  We set
\[
\Ro[r] = \frac{\E[N_{r+1}]}{\E[N_r]}
\]
to be the ratio of the expected number of infections in rank
$r+1$ to the expected number in rank $r$.  This value is
affected by local small-scale structures.  If the network is small, it
is also affected by the finite size of the network, but if the network
is large enough relative to $r$, we expect that the value will be
unaffected by large-scale structure.  In more concrete terms, the
early growth of a disease in a neighbourhood is unaffected by whether that
neighbourhood is part of a city of 100000, 1 million, or 10 million.  As the
disease spreads further, the effect of the finite city size
will be noticeable for the smaller cities first.  If the population is
large enough, the ratio converges before the finite size has any
impact.  We define $\Ro$ mathematically as 
\begin{equation}
\Ro = \lim_{r\to \infty} \lim_{|G|\to\infty} \Ro[r] \, .
\label{eqn:appR0def}
\end{equation}

This definition is similar to that of~\cite{trapman}, who used
\begin{align}
\hat{R}_r &= \E[N_{r+1}]^{1/(r+1)} \nonumber\\
\Ro&=\limsup_{r \to \infty} \limsup_{|G|\to\infty}  \hat{R}_r\, , \label{eqn:trapman}
\end{align}
which is the limit as $r \to \infty$ of the geometric mean of $\Ro[1],
\ldots, \Ro[r-1]$ (assuming the limit exists).  This definition is
more general and will converge in some cases
where~\eqref{eqn:appR0def} does not.  However, if~\eqref{eqn:appR0def}
does converge (and typically we see that it does), then it reaches the
same value, but does so sooner.  So to clearly see
$\Ro$ from~\eqref{eqn:trapman}, we must have a larger network.

\begin{figure}
\includegraphics[width=0.3\columnwidth]{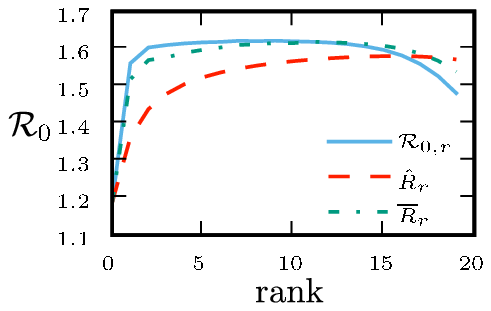}\hfill
\includegraphics[width=0.3\columnwidth]{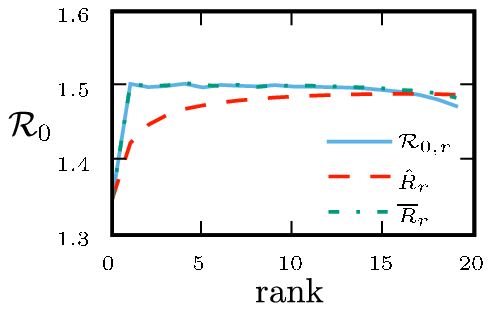}\hfill
\includegraphics[width=0.3\columnwidth]{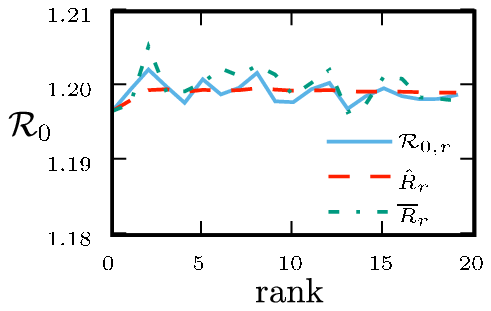}\hfill
\caption{A comparison of the convergence of $\Ro[r]$,
  $\hat{R}_r$, and $\overline{R}_r$ for epidemics in the
  EpiSimS network ($T=0.075$), an unclustered bimodal network ($T=0.3$
  with each node's degree coming either from a Poisson distribution
  peaked at $3$ or a Poisson distribution peaked at $6$), and an
  \erdosrenyi{} network ($T=0.3$, average degree $4$).  The
  calculations used $10^5$ simulations for each network.  Note the difference in vertical scales.}
\label{fig:R0defn_comp}
\end{figure}

Another suitable definition would be
\begin{align}
\overline{R}_r &= \E[N_{r+1}/N_r]\nonumber \\
\Ro&=\lim_{r \to \infty} \lim_{|G| \to \infty} \overline{R}_r \, , \label{eqn:rhat}
\end{align}
where the expectation is taken over realisations with $N_r \neq 0$.
This will tend to require more steps to converge because it
counts small outbreaks equally with large outbreaks, and so outbreaks
which have not yet grown and are dominated by stochastic effects would
be as important to the average as well-established epidemics.  

A comparison of these three definitions of $\Ro$ is shown in
figure~\ref{fig:R0defn_comp}.  They all result in similar values for
$\Ro$.  For a clustered network, equation~\eqref{eqn:appR0def}
converges more quickly.  For large unclustered networks,
$\Ro[r]=\overline{R}_r$ and both converge to $\Ro$ at $r=1$ while
$\hat{R}_r$ takes longer.  In an \erdosrenyi{} network, all three
definitions give $\Ro[r]=\Ro$ for all $r$; only noise due to
insufficient simulations affects the calculation.

To be fully rigorous, the $|G| \to \infty$ limit must be appropriately
defined.  It does not make sense to talk about $|G| \to \infty$ for a
given network, and we cannot simply add nodes to the pre-existing
network.  We must take a sequence of networks in such a way that the
small-scale structure is preserved, and as the network size grows, the
size of the preserved structure increases.

To make this rigorous, we follow~\cite{miller:bounds}.  Take a
sequence of finite networks $G_n$, with $|G_n| \to \infty$ as $n \to
\infty$.  We define $B_r$ to be the network induced on the set of nodes
within distance $r$ of a central node.  The sequence of networks is
taken so that the probability that the structure surrounding a
randomly chosen central node is isomorphic to a given $B_r$ is the
same for all $G_n$ if $n \geq r$.  This means that the small-scale
structure in the different networks is the same, and the size of what
is considered ``small-scale'' increases with $n$.  

We note that although the $|G| \to \infty$ limit may be
well-defined, it is possible that the $r \to \infty$ limit in
\eqref{eqn:appR0def} does not converge.  This may occur because, for
example, growth within a neighbourhood may happen at one rate, while
spread between neighbourhoods in a suburb may happen at another, and
spread between suburbs in a city may happen at yet another.  If the
rate of spread continues to change as the grouping size changes, then
the $r \to \infty$ limit may not exist.  An effect analogous to this
may appear in~\cite{ajelli} which considered disease spread in Italy.
Two distinct growth rates are seen depending on whether the disease is
spreading in the general country or in Rome.

\begin{figure}
\includegraphics[width=0.3\columnwidth]{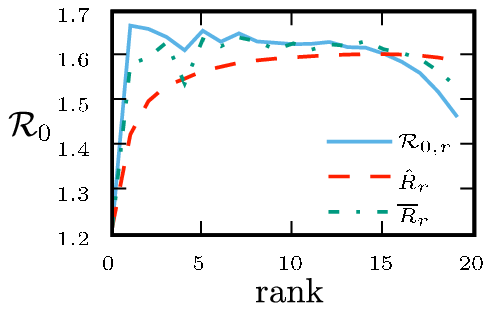}\hfill
\includegraphics[width=0.3\columnwidth]{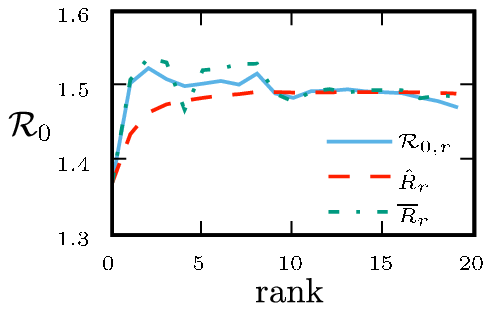}\hfill
\includegraphics[width=0.3\columnwidth]{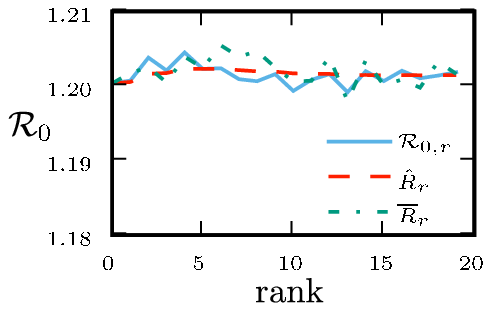}
\caption{A comparison of the convergence of $\Ro[r]$,
  $\hat{R}_r$, and $\overline{R}_r$ for the same networks and conditions as in figure~\ref{fig:R0defn_comp} except using a single EPN for each data point with $10^5$ different index cases within that EPN rather than $10^5$ distinct simulations [actually at submission the first two plots used only $10^3$ index cases, more will be added when calculations complete].}
\label{fig:EPN_R0defn_comp}
\end{figure}

Finally, it is possible to estimate $\Ro$ using a single EPN rather
than multiple simulations.  This is a much faster process, and so it
is possible to do many more simulations to reduce the noise.  However,
because it is chosen from a single EPN, there may be a small
systematic error.  In figure~\ref{fig:EPN_R0defn_comp}, we plot the
values for the same conditions as in figure~\ref{fig:R0defn_comp}.  We
use the same number of index cases and so the noise is comparable.
The value of $\Ro$ is not noticably affected by choosing a single EPN
rather than multiple simulations.  In the calculations in the paper,
we have used a single EPN rather than multiple simulations.

\section{Epidemics in Configuration Model Networks}
\label{app:unclustered_review}
We briefly review previous work for epidemic spread in CM networks.  These are the simplest networks to investigate, and
so the theory has been developed
further than for other networks~\cite{andersson:limit,newman:spread,meyers:directed,kenah:second,miller:heterogeneity,babak:finite,marder}.
See~\cite{miller:bounds,serrano:2} for some discussion
of more arbitrary unclustered networks.\footnote{Perhaps the most
  significant result for non-CM networks is that if
  the higher degree nodes preferentially contact other high degree
  nodes, then the threshold transmissibility for an epidemic is
  reduced.}
We extend the earlier theory by allowing 
independently assigned edge weights.\footnote{If
  edge weights are not assigned independently, then infection along
  different edges is not independent, and the methods of this section
  do not apply.}

\subsubsection{The basic reproductive ratio}
Early in the spread of an infectious disease on a CM
network, the probability of a node becoming infected is proportional
to its degree, and so the pdf for the degree of infected nodes is
$kP(k)/\ave{k}$.  We choose an infected node $u$ with degree $k$ uniformly from
nodes of rank $r$.  If the network is large enough that
we can ignore short cycles, then all of $u$'s neighbours are
susceptible except the node which infected $u$.  Thus $u$ may infect
up to $k-1$ neighbours.  The probability $T_{out}(u)$ that $u$ will
infect a randomly chosen neighbour is chosen from $Q_{out}(T_{out})$,
and so the probability $u$ infects exactly $j \leq k-1$ neighbours is
$\binom{k-1}{j} T_{out}(u)^j [1-T_{out}(u)]^{k-1-j}$.  Integrating
this over possible values of $T_{out}$ and summing over $k$ and $j$,
we find that for $r>0$ the rank reproductive ratio is
\[
\Ro[r]  = \frac{1}{\ave{k}} \sum_{k=1}^\infty \left(k P(k) \sum_{j=0}^{k-1}
j \int \binom{k-1}{j}T_{out}^j (1-T_{out})^{k-1-j} P(T_{out}) dT_{out} \right)
 = \ave{T} \frac{\ave{k^2-k}}{\ave{k}} \, ,
\]
and so 
\begin{equation}
\Ro=\ave{T} \frac{\ave{k^2-k}}{\ave{k}}
\label{eqn:R0unclustered}
\end{equation}
Thus we find that for CM networks\footnote{Unless the degree
  distribution satisfies $\ave{k^2-k}=\ave{k}^2$.  The best-known such
  networks are \erdosrenyi{} networks which have a Poisson degree
  distribution in the limit of large network size.}  $\Ro \neq \Ro[0]
= \ave{T}\ave{k}$.

\subsubsection{Probability and size}

We look for the probability that a single infected node causes a chain
of infections leading to an epidemic.  Because interchanging edge
direction in an EPN interchanges $\PE$ and $\A$, we may focus on
calculating $\PE$.  Equivalent techniques replacing $T_{out}$ by
$T_{in}$ below give $\A$.  Our analysis is performed in the infinite
network limit.

We set $f$ to be the probability a randomly chosen index case does not
start an epidemic.  We find
\[
f = \sum_k \left( P(k) \int_{T_{out}} [1-T_{out} + T_{out} h]^k P(T_{out}) dT_{out} \right ) \, ,
\]
where $h$ is the probability a randomly chosen secondary case does not start an epidemic.  The value of $h$ satisfies the recurrence relation
\[
h = \frac{1}{\ave{k}} \sum_k \left( kP(k) \int_{T_{out}} [ 1-T_{out} + T_{out}h]^{k-1} P(T_{out}) \, dT_{out} \right)\, .
\]
If $\Ro <1$, the trivial solution $f=h=1$ is the only solution.  For
$\Ro>1$ an additional solution appears and is the physically relevant
root.  From this we can calculate $\PE = 1-f$.

Note that $\PE$ depends on the distribution of $T_{out}$, but is not
affected by the distribution of $T_{in}$.  Similarly, $\A$ depends on
the distribution of $T_{in}$ but is not affected by the distribution
of $T_{out}$.  This result holds for unclustered, but not for
clustered, networks.


\subsubsection{Summary}
We have shown that for CM networks, $\Ro = \ave{T}
\ave{k^2-k}/\ave{k}$.  In particular it depends only on the network
properties and the average transmissibility.  In contrast, the
probability $\PE$ and size $\A$ are affected by the details of the distribution.
Intuitively, this is easy to understand.  For example, if we consider
$\A$ in populations with varying $T_{in}$, at early
times the rate of growth is governed by the average number of new
infections created, which depends on the average transmissibility.
However, a disproportionate number of highly susceptible nodes are
infected, and so the average $T_{in}$ of remaining nodes drops.  By
the end of the epidemic nodes are much harder to infect than they
would have been if all were equally susceptible initially, and so the
epidemic infects fewer people.

A consequence of this is that we cannot predict $\A$ based only on the
early growth rate.  Although this is frequently done (see for
example~\cite{ma} and references therein), these calculations usually
assume that the population is homogeneously susceptible, which is not
always the case, particularly when a vaccine or previous exposure to
similar diseases exists.

\section{The EpiSimS Network}
\label{app:episims}
\begin{figure}
\begin{center}
\includegraphics[width=0.48\columnwidth]{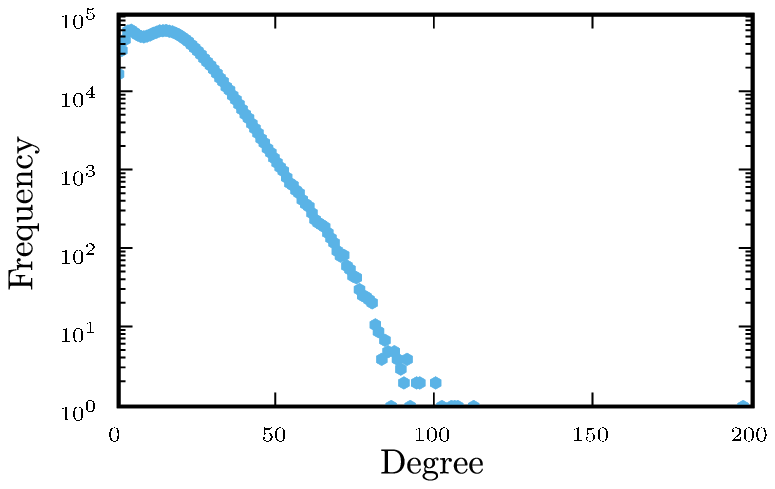}
\hfill
\includegraphics[width=0.48\columnwidth]{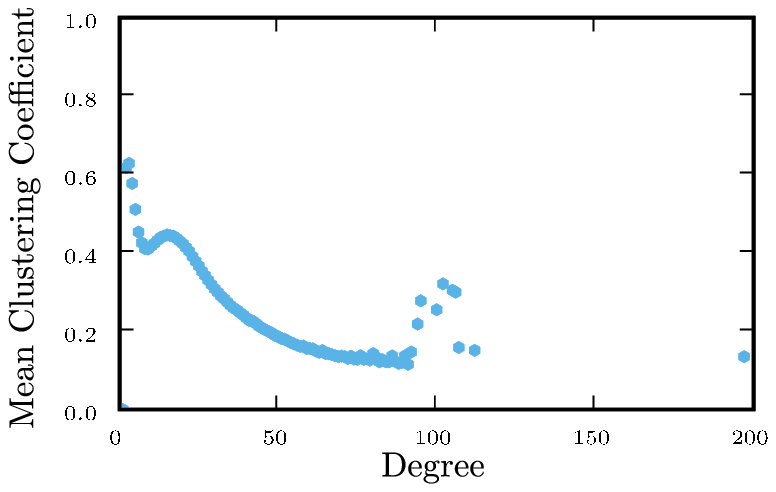}\\
\includegraphics[width=0.48\columnwidth]{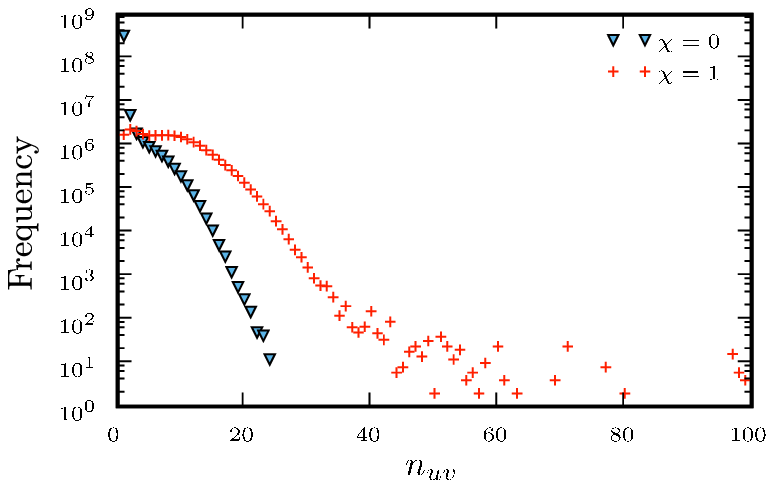}\hfill
\includegraphics[width=0.48\columnwidth]{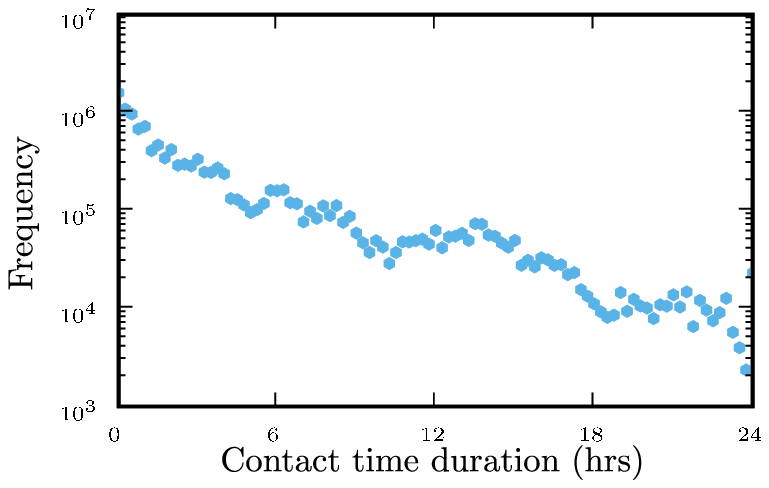}
\end{center}
\caption{Properties of the EpiSimS network.  For the final plot,
  contact times are binned in quarter hour increments, but exact
  values were used in calculations.}
\label{fig:episims_props}
\end{figure}

We consider a network produced by EpiSimS for Portland,
Oregon~\cite{valle:episims,eubanks:episims,barrett}.  This simulation
uses Census data, road structure, building locations, and population
surveys to construct a virtual population that travels through the
city.  From the activity of individuals in the simulation, we may reconstruct
who was in contact with whom and for how long.

There are $1615860$ nodes in the network, of which $1591010$ are in
the giant component.  The average degree is approximately $16$, and
the average squared degree is approximately $359$.  The degree
distribution has an exponential tail, and clustering is concentrated
in the low-degree nodes.  For our approximations of $\Ro$, we also
need information about length $2$ paths.  We calculate the number of
pairs of nodes with each value of $n_{uv}$ for which $\chi_{uv}=0$
and $\chi_{uv}=1$.  Large values of $n_{uv}$ are more frequent when
$\chi_{uv}=1$.  The distribution of edge weights is fairly broad.
Many contacts are very short, but the number of long contacts is not
negligible.


\end{document}